\documentclass[
	aps,
	prd, 
	twocolumn, 
	a4paper, 
	10pt, 
	amsmath,
	amssymb, 
	preprintnumbers,  
	tightenlines, 
	notitlepage,
	floatfix,
	nofootinbib,
	longbibliography
	]{revtex4-1}\pdfoutput=1
\usepackage[utf8]{inputenc}
\usepackage[english]{babel}
\usepackage{bbm}
\usepackage{graphicx}
\usepackage{hyperref}
\usepackage{wasysym}
\usepackage{amsthm}
\usepackage{dcolumn}
	\newcolumntype{.}{D{.}{.}{13}}
	\newcolumntype{d}[1]{D{.}{.}{#1}}
\usepackage{color}	
\usepackage[all,cmtip]{xy}
\usepackage{tensor}

\graphicspath{{pics/}}

\allowdisplaybreaks[1]

\newcommand{\ZZ}{\mathbbm{Z}}					

\newcommand{\abs}[1]{\lvert#1\rvert}			
\newcommand{\norm}[1]{\lVert#1\rVert}			
\newcommand{\avg}[1]{\left\langle #1 \right\rangle }		

\newcommand{\cbB}[1]{\Big\{#1\Big\} }			
\newcommand{\hh}[1]{\left(#1\right) }			
\newcommand{\bh}[1]{\bigl(#1\bigr) }			
\newcommand{\Bh}[1]{\Bigr(#1\Bigr) }			

\newcommand{\id}[1]{\operatorname{d}\!#1}												
\renewcommand{\d}[2]{\frac{\operatorname{d}\!#1}{\operatorname{d}\!#2}}					
\newcommand{\dd}[2]{\frac{\operatorname{d}^2\!#1}{\operatorname{d}\!#2^2}}				
\newcommand{\pd}[2]{\frac{\operatorname{\partial}\!#1}{\operatorname{\partial}\!#2}}	
\newcommand{\CD}[1]{\nabla_{#1}}

\newcommand{\ii}{i}								
\newcommand{\ee}{e}								

\newcommand{\pt}{\tau}							
\newcommand{\mt}{\lambda}						

\newcommand{\nE}{\mathcal{E}}								
\newcommand{\nL}{\mathcal{L}}							

\newcommand{\rmin}{r_\mathrm{min}}
\newcommand{\rmax}{r_\mathrm{max}}

\newcommand{\mr}{\eta}

\DeclareMathOperator{\bigO}{\mathcal{O}}					

\newcommand{\tet}[2]{e_{#1}^{#2}}				
\newcommand{\itet}[2]{e^{#1}_{#2}}				

\newcommand{\R}[3][]{\,{_{#2}R_{#3}^{\mathrm{#1}}}}		
\newcommand{\Rb}[3][]{\,{_{#2}\bar{R}_{#3}^{\mathrm{#1}}}}		
\newcommand{\SWSH}[3][]{\,{_{#2}S_{#3}^{\mathrm{#1}}}}	
\newcommand{\Y}[3][]{\,{_{#2}Y_{#3}^{\mathrm{#1}}}}		
\newcommand{\T}[3][]{\,{_{#2}T_{#3}^{\mathrm{#1}}}}		

\newcommand{\Sb}[4]{({_{#1}b_{#2}})^{#3}_{#4}}			
\newcommand{\YA}[4]{\tensor*[_{#1\hspace{3pt}}^{#2}]{\hspace{-3pt}\mathcal{A}}{^{#3}_{#4}}
}			
\newcommand{\YB}[4]{{^{#1#2}\!\mathcal{B}^{#3}_{#4}}}			

\newcommand{\TEV}[2]{\,{_{#1}\lambdabar_{#2}}}			
\newcommand{\SEV}[2]{\,{_{#1}A_{#2}}}					

\newcommand{\pI}{+}						
\newcommand{\pH}{-}						
\newcommand{\pIH}{\pm}					

\newcommand{\MROP}[1]{\hat{\mathcal{H}}_{#1}}	

\newcommand{\MC}{\mathcal{C}}					

\newcommand{\SWL}[1]{\bar\eth_{#1}}				
\newcommand{\TP}{\epsilon}						

\newcommand{\spl}{\mathfrak{l}}					

\newcommand{\lmax}{l_\mathrm{max}}				
\newcommand{\llmax}{\spl_{\mathrm{max}}}		

\newcommand{\ret}{\mathrm{Ret}}

\newcommand{\reg}{\mathrm{R}}
\newcommand{\sing}{\mathrm{S}}

\newcommand{\Lor}{\mathrm{Lor}}
\newcommand{\Rad}{\mathrm{Rad}}
\newcommand{\Avg}{\mathrm{Avg}}

\DeclareMathOperator{\Dop}{\hat{D}}
\DeclareMathOperator{\Delop}{\hat{\Delta}}
\DeclareMathOperator{\delop}{\hat{\delta}}
\DeclareMathOperator{\delopbar}{\hat{\bar{\delta}}}

\newcommand{\Fext}{\mathcal{F}} 						

\newcommand{\umin}{\mathord{\scalebox{0.75}[1]{\(-\)}}}

\newcommand\Ts{\rule{0pt}{2.6ex}}       
\newcommand\Bs{\rule[-1.2ex]{0pt}{0pt}} 

\newcommand{\solarmass}{M_{\astrosun}}
\begin{document}

\title{Gravitational self-force on eccentric equatorial orbits around a Kerr black hole}

\author{Maarten \surname{van de Meent}}
\email{M.vandeMeent@soton.ac.uk}
\affiliation{Mathematical Sciences, University of Southampton, Southampton, SO17 1BJ, United Kingdom}

\date{\today}
\begin{abstract}
This paper presents the first calculation of the gravitational self-force on a small compact object on an eccentric equatorial orbit around a Kerr black hole to first order in the mass-ratio. That is the pointwise correction to the object's equations of motion (both conservative and dissipative) due to its own gravitational field, which is treated as a linear perturbation to the background Kerr spacetime generated by the much larger spinning black hole. The calculation builds on recent advances on constructing the local metric and self-force from solutions of the Teukolsky equation, which led to the calculation of the Detweiler-Barack-Sago redshift invariant on eccentric equatorial orbits around a Kerr black hole in a previous paper.

After deriving the necessary expression to obtain the self-force from the Weyl scalar $\psi_4$, we perform several consistency checks of the method and numerical implementation, including a check of the balance law relating the orbital average of the self-force to the average flux of energy and angular momentum out of the system. Particular attention is paid to the pointwise convergence properties of the sum over frequency modes in our method, identifying a systematic inherent loss of precision that any frequency domain calculation of the self-force on eccentric orbits must overcome.

\end{abstract} 

\maketitle
\setlength{\parindent}{0pt} 
\setlength{\parskip}{6pt}

\section{Introduction}
With LIGO's detection of the first gravitational wave event GW150914 \cite{Abbott:2016blz} the era of gravitational wave astronomy has begun in earnest. This enterprise crucially depends on the availability of accurate gravitational wave templates to extract physical information from the gravitational wave signal. In the case of GW150914, these templates were provided by a combination of numerical relativity (NR), post-Newtonian (PN), and effective one-body (EOB) methods.

These methods work well for binaries consisting of two compact objects with masses $m$ and $M$, whose ratio ${\mr=m/M}$ is comparable to 1 (as was the case for the source of GW150914). However, these methods struggle as the mass-ratio $\mr$ becomes small. The large disparity in length scales set by the gravitational radii of the objects in this situation makes full NR simulations unfeasible. Moreover, systems with a small mass-ratio spend a large number ($\sim \eta^{-1}$) of orbits in the strong field regime where PN approximations become inaccurate. In principle EOB methods should be able to cover this regime; however, current implementations calibrated using NR and PN data are not guaranteed to be accurate.

Nonetheless, the small mass-ratio regime is of great physical interest. Historically, this interest has been much motivated by the prospect of observing extreme mass ratio inspirals or EMRIs -- compact binaries consisting of a stellar mass compact object orbiting a supermassive black hole -- with a space-based gravitational wave observatory, like ESA's planned eLISA mission (currently scheduled for launch in the mid 2030s). EMRIs are thought to occur regularly in most galactic nuclei and can be observed with eLISA up to cosmological distances. Observations would allow accurate ($\sim 10^{-5}$) measurement of the system's properties including orbital parameters, mass, spin, and (luminosity) distance \cite{Barack:2003fp}. Alternatively, the observations can be used to test the hypothesis that the geometry of the host black hole is described by the Kerr geometry to high accuracy \cite{Barack:2006pq}.

The surprisingly large black hole masses in the LIGO observations (GW150914 had $m=29 \solarmass$ and $M=36 \solarmass$) further raise the possibility of the occurrence of intermediate mass-ratio inspirals (IMRIs) consisting of a stellar mass object orbiting a $\sim 100 \solarmass$ object. Even a $1.4\solarmass$ neutron star orbiting a $36 \solarmass$ black hole would be challenging for current NR methods.

Study of the small-ratio regime is of further interest for the synergy with other methods for modelling black hole binaries that can be obtained by comparing results in overlapping regimes of validity \cite{LeTiec:2011dp,LeTiec:2011bk}. In particular, the last couple of years have seen some much useful synergy in using small mass-ratio data to refine EOB models\cite{Barausse:2011dq,Akcay:2012ea,Akcay:2015pza,Bini:2013zaa,Bini:2014nfa,Bini:2014zxa,Bini:2015bfb,Bini:2015mza,Bini:2016qtx}, and self-force calculations have been essential in fixing ambiguities in the recent derivation of the 4PN equations of motion for non-spinning black hole binaries \cite{Damour:2014jta,Jaranowski:2015lha,Bernard:2015njp}.

Small mass-ratio binaries can be modelled by treating the small mass-ratio $\eta$ as a perturbative parameter. At zeroth order in $\eta$, the smaller mass $m$ becomes a test particle and will follow a geodesic of the Kerr spacetime generated by the larger mass $M$, which can be solved analytically \cite{Fujita:2009us,Hackmann:2008zz,Hackmann:2010tqa,Hackmann:2010zz}. At the next order in perturbation theory, the corrections to the motion of the smaller object can be summarized by an effective force term in the geodesic equation, the \emph{gravitational self-force (GSF)}. 

The first formal expressions for the GSF were introduced by Mino, Sasaki, and Tanaka \cite{Mino:1996nk} and Quinn and Wald \cite{Quinn:1996am}, two decades ago. In the years since, their formalism has been further refined (see \cite{Poisson:2011nh,Pound:2015tma} for reviews and references) increasing both mathematical rigour and conceptual clarity. According to this formalism, the (first-order) GSF can be calculated by finding the linear metric perturbation sourced by a point particle following a background geodesic and isolating a particular finite contribution at the particle's location. A practical procedure (known as \emph{mode sum regularization}) for determining this finite piece was introduced by Barack and Ori \cite{Barack:2001gx,Barack:2001ph,Barack:2002bt} around the turn of the millennium. 

This method has been implemented numerically for particles on increasingly complicated orbits. The first calculations were done in 2002 for a particle falling radially into a Schwarzschild black hole \cite{Barack:2002ku}. Circular orbits followed in 2007 \cite{Barack:2007tm} and the GSF on eccentric orbits was first calculated in 2009 \cite{Barack:2009ey}. These calculations relied on the fact that the linearized Einstein equation on a Schwarzschild background can be decoupled into separate $1+1$-dimensional partial differential equations for each spherical harmonic mode, which can be solved 1-by-1 in the time domain. Further computational efficiency can be gained by Fourier transforming to the frequency domain, leading to a system of decoupled linear ordinary differential equations \cite{Detweiler:2008ft,Hopper:2010uv,Akcay:2010dx,Akcay:2013wfa,Osburn:2014hoa}.

Extending these calculations to the scenario where the larger black hole has spin and produces a Kerr spacetime has proven much more difficult. One of the main issues is that in Kerr spacetime the linearized Einstein equation cannot be solved by separation of variables. Several approaches to circumvent this problem have been explored.

Dolan and Barack \cite{Dolan:2010mt,Dolan:2011dx,Dolan:2012jg} have used the axisymmetry of the background to separate out the angular $\phi$ dependence from the Lorenz gauge field equations and then numerically solved the remaining $2+1$ dimensional time domain equations. Besides the obvious numerical costs, this method is troubled by some numerical instabilities. Nonetheless, these problems have been overcome to calculate the self-force on circular equatorial orbits \cite{Dolanunpublished}. 

Another approach builds on the fact that the Weyl scalars $\psi_0$ and $\psi_4$ in Kerr spacetime satisfy the Teukolsky equation \cite{Teukolsky:1972my,Teukolsky:1973ha}, which is separable in the frequency domain. Moreover, a key result of Wald \cite{Wald:1973} shows that these Weyl scalars contain almost all gauge invariant information about the metric perturbation up to a global perturbation of the mass and angular momentum of the Kerr background. Chrzanowski, Cohen, and Kegeles \cite{Chrzanowski:1975wv,Cohen:1974cm,Kegeles:1979an} have provided an explicit method for reconstructing the metric perturbation in radiation gauge from either $\psi_0$ or $\psi_4$.

The group of Friedman in Milwaukee has pioneered the use of this construction to calculate the gravitational self-force \cite{Keidl:2006wk,Keidl:2010pm,Shah:2010bi,Shah:2012gu}. There have been two longstanding issues with this approach. The first is that metric in the radiation gauge is known to be highly irregular in the presence of matter sources \cite{Ori:2002uv}. Not only does a point particle create a divergence at its location, it is also invariably accompanied by a string-like singularity extending from the particle to infinity and/or the black hole horizon. This posed a problem since the derivations of the self-force assume the metric perturbation to have a singularity structure similar to the Lorenz gauge. A detailed analysis of this problem by Pound et al. \cite{Pound:2013faa} has however shown that the self-force can indeed be calculated in particular choices of the radiation gauge.

A second problem is the missing mass and angular momentum perturbations. On Schwarzschild backgrounds, Birkhof's theorem implies a particularly simple solution: Outside of the particle's orbit the mass and angular momentum perturbations are given by the energy and angular momentum of the particle and vanish inside the orbit. However, no such straightforward argument appears to be available on Kerr backgrounds. Nonetheless, Merlin et al. proved \cite{completion,MerlinThesis} that imposing analyticity of certain gauge invariant fields constructed from the metric away from the particle implies the same simple result remains true in Kerr spacetimes for all bound equatorial orbits.

Assuming the above two results (without proof) Friedman's group successfully implemented above method to calculate the Detweiler redshift invariant \cite{Detweiler:2008ft} for circular equatorial orbits first in Schwarzschild spacetime \cite{Shah:2010bi} and later in Kerr \cite{Shah:2012gu}. More recently, the author together with Shah used the newly available missing pieces to implement a numerical calculation of the (generalized) redshift invariant \cite{vandeMeent:2015lxa} for eccentric equatorial orbits. The main goal of this paper is to provide the first calculation of the full self-force on eccentric equatorial orbits in Kerr spacetime.

The plan for this paper is as follows. Section~\ref{sec:prelim} reviews the preliminaries of the self-force formalism needed for our calculations. We then continue to discuss the details of our method in Sec.~\ref{sec:method}. In particular, we derive the explicit expressions needed to calculate the gravitational self-force from a given frequency domain solution of the Teukolsky equation for $\psi_4$. In Sec.~\ref{sec:implementation}, we provide some details of the numerical implementation of our method. Section \ref{sec:results} presents a number of consistency checks of our method and numerical implementation. Finally, we conclude with a discussion of our results and conclusions in Sec.~\ref{sec:discussion}.

\subsection{Conventions}
This paper uses an overall metric signature of $(-+++)$; for further sign conventions regarding the definitions of other quantities such as the Weyl curvature scalars we refer to Appendix \ref{app:A}. We further work in geometrized units such that $(c=G=M=1)$.

\section{Premliminaries}\label{sec:prelim}

\subsection{Gravitational self-force}
Suppose we have a binary system consisting of two objects with masses $m$ and $M$, which are both compact in the sense that their size is of the order of their respective gravitational length scales set by their masses. The goal of the self-force programme is to describe the dynamics of such a binary using perturbation theory with the mass-ratio $\mr=m/M$ as a small parameter. At zeroth order in $\mr$, the smaller object acts as a test mass in the geometry generated by the larger mass $M$, with its trajectory $x_0^\mu(\pt)$ obeying the geodesic equation,
\begin{equation}
\dd{x_0^\mu}{\pt}+ \Gamma^{\mu}_{\alpha\beta}\d{x_0^\alpha}{\pt}\d{x_0^\beta}{\pt} =0,
\end{equation}
where $\pt$ is proper time and $\Gamma^{\mu}_{\alpha\beta}$ the usual Christoffel symbols. At first order in $\mr$, the metric generated by the binary can be split as
\begin{equation}\label{eq:metsplit}
 g_{\mu\nu} + \mr h_{\mu\nu},
\end{equation}
where $g_{\mu\nu}$ is the background Kerr geometry generated by $M$, and $h_{\mu\nu}$ is some linear perturbation generated by $m$. Clearly, $h_{\mu\nu}$ should satisfy the linearized Einstein equation; however, it is not immediately clear what should be used as a source term. Moreover, we would like to describe the motion of $m$ by some effective force correction to the geodesic equation,
\begin{equation}\label{eq:gdGSF}
\mr^2 F^\mu[h] \equiv m\hh{\dd{x_0^\mu}{\pt}+ \Gamma^{\mu}_{\alpha\beta}\d{x_0^\alpha}{\pt}\d{x_0^\beta}{\pt}},
\end{equation}
the \emph{gravitational self-force} or GSF. However, it is far from obvious how to obtain $F^\mu$. For starters, given that $m$ has some physical extent it is not even obvious how to define $m$'s position $x^{\mu}(\pt)$. These questions are most rigorously addressed using a multiscale expansion as described in the reviews \cite{Poisson:2011nh,Pound:2015tma}. We will not describe the details here, but the general gist is to describe general solutions to the Einstein equation in a small region near $m$ where the background metric $g$ is approximately flat, and in a far region where $h$ is properly small, and matching the solutions in an intermediate region where both approximations hold simultaneously. The upshot is that at linear order in $\mr$, the appropriate source for $h_{\mu\nu}$ is a point particle of mass $m$ following a trajectory $x^{\mu}(\pt)$ defined by $m$'s centre-of-mass.

Furthermore, 
 $x^{\mu}(\pt)$ satisfies the geodesic equation in the spacetime $g_{\mu\nu}+\mr h_{\mu\nu}^\reg$, where $h_{\mu\nu}^\reg$ is a certain smooth part of $h_{\mu\nu}$ first identified in \cite{Detweiler:2002mi}. The GSF $F^\mu$ is then given by the MiSaTaQuWa \cite{Mino:1996nk,Quinn:1996am} equation,
\begin{equation}\label{eq:GSFdef}
F^\mu(\tau) =  P^{\mu\alpha\beta\gamma}\CD{\alpha}h^\reg_{\beta\gamma}(x_0(\pt)),
\end{equation}
with
\begin{equation}
P^{\mu\alpha\beta\gamma} \equiv 
	\frac{1}{2}\hh{
		g^{\mu\alpha}u^\beta u^\gamma
		-2g^{\mu\beta}{u}^\alpha{u}^\gamma
		-{u}^\mu{u}^\alpha{u}^\beta{u}^\gamma
	},
\end{equation}
where $u^\mu$ is the four-velocity $\d{x^\mu_0}{\pt}$ (in the background spacetime). If $m$ has non-zero intrinsic angular momentum, this is supplemented by a term depending on the object's spin dipole moment as found by Papapetrou \cite{Papapetrou:1951pa}. This term depends only on the background metric and will not be considered further in this paper.

Although in this paper we will only be considering the order $\mr$ corrections to the dynamics of the binary, it is worth mentioning that the same picture extends to general orders in perturbation theory \cite{Pound:2015tma}. In general, at any order in perturbation theory $m$ will follow the trajectory of a point particle in some effective metric, supplemented by corrections due to a finite number of multipole moments.

The perturbative procedure above intimately depends on the chosen split in \eqref{eq:metsplit} between a background $g$ and perturbation $h$, which is not unique. One could chose a different \emph{gauge} by considering coordinates $\tilde{x}^\mu$ that differ from $x^\mu$ by a small amount $\mr \xi^\mu$, and ascribing the resulting shift in the components of $g$ to the perturbation $h$. Performing the perturbative procedure above in this new gauge leads to a self-force $ \tilde{F}^\mu$ that is changed by
\begin{equation}
 \tilde{F}^\mu - F^\mu = -\hh{g^{\mu\alpha}+u^\mu u^\alpha}\CD{u}^2\xi_\alpha-\tensor{R}{^\mu_\alpha_\beta_\gamma}u^\alpha\xi^\beta u^\gamma.
\end{equation}
In practical calculations, this gauge freedom is fixed by imposing a gauge condition on $h$. Traditionally most self-force calculations have been done in the \emph{Lorenz gauge} defined by
\begin{equation}\label{eq:lorgauge}
	\CD{\alpha}\hh{h^{\alpha\mu}-\frac{1}{2}g^{\alpha\mu}g^{\beta\gamma}h_{\beta\gamma}}=0.
\end{equation}
The method described in this paper produces the self-force in the \emph{outgoing radiation gauge}, which in vacuum regions is defined through the conditions
\begin{alignat}{3}
h_{2a}&= \tet{2}{\mu}\tet{a}{\nu}h_{\mu\nu} &&=0,\\
h_{34}&= \tet{3}{\mu}\tet{4}{\nu}h_{\mu\nu} &&=0,
\end{alignat}
where the $\tet{a}{\mu}$ form are null tetrad (see Appendix~\ref{app:A} for details).

\subsection{Mode sum regularization}
One of the main challenges in any practical calculation of the GSF is determining the regular part of the metric perturbation, $h_{\mu\nu}^\reg$. Over the last two decades various schemes have been introduced (see \cite{Barack:2009ux} and \cite{Wardell:2015kea} for reviews). We here adopt the so-called \emph{mode sum regularization} scheme \cite{Barack:2001gx,Barack:2001ph,Barack:2002bt}, which we will review presently.

This method starts from the observation that the regular field $h_{\mu\nu}^\reg$ can be expressed as the difference
\begin{equation}\label{eq:hRdiff}
h_{\mu\nu}^\reg = h_{\mu\nu}^\ret - h_{\mu\nu}^\sing
\end{equation}
between the retarded field $h_{\mu\nu}^\ret$ (i.e. the solution of the linearized Einstein equation with a point particle source and retarded boundary conditions), and the Detweiler-Whiting singular field $h_{\mu\nu}^\sing$, which solves the same linearized Einstein equation but is constructed such that it does not contribute to the self-force.

Unfortunately, both terms on the right hand side of Eq.~\eqref{eq:hRdiff} diverge at the location of the particle. Consequently, this subtraction makes sense everywhere except at the location where we need $h_{\mu\nu}^\reg$ to calculate the GSF through \eqref{eq:GSFdef}. We thus need a regularization mechanism. The chosen mechanism is to decompose all fields in spherical harmonic ``$l$-modes''. For any field $f(x)$, its $l$-modes are defined by
\begin{equation}\label{eq:lmodedef}
f_l(x) \equiv \sum_{m=-l}^l
\Bh{\int_{S^2}\hspace{-8pt}\id{\Omega} f\bar{Y}_{lm}}Y_{lm}(z,\phi),
\end{equation}
where the integral is performed over a sphere of constant $t$ and $r$. The key observation is that these $l$-modes have a finite (although possibly directionally dependent) limit at the particle location $x_0$.

In principle, the decomposition into $l$-modes could be done at the level of the metric perturbation and its derivatives. However, following \cite{Barack:2001gx,Barack:2001ph,Barack:2002bt}, we promote the self-force to a field $\Fext$, and decompose this \emph{extended} field into $l$-modes. Promotion of the self-force to a field requires extending \eqref{eq:GSFdef}, which was defined only at the particle worldline, to a field equation. At the very minimum this requires extending the four-velocity $u$ to a field, but more generally any field equation that reduces to \eqref{eq:GSFdef} on the worldline can be used. We will follow \cite{Barack:2009ux} and choose to extend \eqref{eq:GSFdef} to a field by promoting the four-velocity $u^\mu$ to a field $\hat{u}^\mu$ defined to be constant on each constant $t$-slice and take its natural value at the worldline $x_0$.

With this choice of extension it is possible to obtain a Laurent expansion of $\Fext^\mu_{\sing}$ in the Lorenz gauge \cite{Barack:2002bt,Barack:2009ux,Mino:2001mq}, and in turn the large $l$ behaviour of its $l$-modes,
\begin{equation}\label{eq:ABCdef}
\begin{split}
F^{\mu,\pm}_{l,\sing} &\equiv\lim_{x\to x_0^\pm} \Fext^\mu_{\sing,l}\\
 &=  \pm L A^{\mu}_\Lor + B^{\mu}_\Lor +\frac{C^{\mu}_\Lor}{L} + \bigO(L^{-2}), 
\end{split}
\end{equation}
with $L=l+1/2$, and where the $\pm$ sign depends on from which radial direction $x_0$ was approached. It is further possible to show that in this extension,
\begin{equation}\label{eq:Ddef}
D^{\mu}_\Lor \equiv \sum_l F^{\mu,\pm}_{l,\sing} \mp L A^{\mu}_\Lor - B^{\mu}_\Lor -\frac{C^{\mu}_\Lor}{L}=0.
\end{equation}
The quantities $A^{\mu}_\Lor$, $B^{\mu}_\Lor$, $C^{\mu}_\Lor$, and $D^{\mu}_\Lor$ are collectively known as \emph{regularization parameters}. If one can calculate the self-force $l$-modes of the Lorenz gauge retard field in the same extension, then one can calculate the actual self-force from the difference of the retarded and singular field $l$-modes using the mode-sum formula,
\begin{equation}\label{eq:modesum}
F^\mu = \Bh{\sum_l F^{\mu,\pm}_{l,\Lor} \mp L A^{\mu}_\Lor - B^{\mu}_\Lor -\frac{C^{\mu}_\Lor}{L}}-D^{\mu}_\Lor.
\end{equation}

However, in this paper, we obtain the self-force not in the Lorenz gauge, but in the outgoing radiation gauge (ORG). This introduces complications because in the presence of matter the metric perturbation in this gauge cannot be made regular everywhere in the vacuum part of the spacetime \cite{Ori:2002uv}. With a point particle source there will be a stringlike singularity in $h$ extending from the particle towards infinity and/or the black hole horizon. 

In \cite{Pound:2013faa}, the effect of these string singularities on the calculation of the self-force was studied in detail. Several approaches to calculating the self-force from radiation gauge data are offered. We here follow their ``no-string'' approach. If regularity of $h$ at infinity is imposed, the metric is obtained in a variant of the ORG that has a half-string singularity extending from the particle to the black hole horizon. Conversely, imposing regularity on the horizon produces a half-string singularity extending from the particle to infinity. A metric perturbation with no string singularities can be constructed by taking the regular halves of two half-string solutions and glueing them together along a timelike hypersurface containing the particle trajectory. This comes at the price of introducing a discontinuity in the metric perturbation along this hypersurface.

It was further shown in \cite{Pound:2013faa} that the self-force formalism can be extended to apply to irregular metric perturbations with half-string singularities. In fact, the regularization parameters $A^\mu$, $B^\mu$, and $C^\mu$ appearing in the mode-sum formula take the same values in the "half-string" radiation gauge as they do in the Lorenz gauge if the extension of the self-force is kept the same. Unfortunately, the regularization parameter $D^\mu$ does receive a finite correction in these gauges. However, it is observed that this correction differs between the two half-string gauges only in sign. Consequently, if one calculates the GSF in the discontinuous no-string gauge where it is simply given by the average of the two half-string solutions, the Lorenz gauge values can be used for all regularization parameters. Accordingly, \cite{Pound:2013faa} obtain a modified mode-sum formula taking radiation gauge data as its input and using the Lorenz gauge regularization parameters,
\begin{equation}\label{eq:modesumavg}
F^\mu = \Bh{\sum_l \frac{F^{\mu,+}_{l,\Rad}+F^{\mu,-}_{l,\Rad}}{2} - B^{\mu}_\Lor -\frac{C^{\mu}_\Lor}{L}}-D^{\mu}_\Lor.
\end{equation}

\subsection{Eccentric geodesics}
As noted above, at order zero in the mass-ratio the motion of the smaller body is described by a geodesic in the Kerr spacetime generated by the larger body. As shown by Carter \cite{Carter:1968rr}, the geodesic equation can be reduced to a set of first-order equations,
\begin{align}
\Bh{\d{r}{\pt}}^2 &=\frac{ R(r)}{\Sigma(r,z)^2}, \\
\Bh{\d{z}{\pt}}^2 &= \frac{Z(z)}{\Sigma(r,z)^2},  \\
\d{\phi}{\pt} &=\frac{\Phi_r(r)+\Phi_z(z)}{\Sigma(r,z)},  \\
\d{t}{\pt} &= \frac{\mathfrak{T}_r(r)+\mathfrak{T}_z(z)}{\Sigma(r,z)},  
\end{align}
where $R$, $Z$, $\Phi_r$, $\Phi_z$, $\mathfrak{T}_r$, and $\mathfrak{T}_z$ are known functions of the Boyer-Lindquist coordinates $r$ and $z=\cos\theta$ (see e.g. \cite{Fujita:2009bp}), and $\Sigma$ is defined below. This set of equations can easily be separated by changing to a convenient time variable $\mt$ to parametrize the orbit,
\begin{equation}
 \d{\pt}{\mt}=\Sigma(r,z) =  r^2+a^2 z^2.
\end{equation}
This time variable $\mt$ is commonly referred to as ``Mino time''. With this choice of time parameter, the radial ($r$) and polar ($z$) motion satisfy separate differential equations. For bound geodesics, each motion has its own frequency  $\Upsilon_r$ and $\Upsilon_z$. The position along a bound geodesic is therefore uniquely determined by two phases $q_r = \Upsilon_r\mt$ and $q_z = \Upsilon_z\mt$. Complete analytic solutions of the geodesic equations as functions of $q_r$ and $q_z$ were given by \cite{Fujita:2009bp}.

In this paper we restrict ourselves to equatorial orbits with $z=0$ (we can thus ignore the polar phase $q_z$). Up to shifts in $t$, $\phi$, and  radial phase $q_r$, bound equatorial geodesics are uniquely determined by two parameters. One could for example use the (specific) energy $\nE$ and angular momentum $\nL$ of the orbit. However, it is convenient for us to use the semilatus rectum $p$ and eccentricity $e$, defined by
\begin{align}
\rmin &= \frac{p}{1+e},\\
\rmax &= \frac{p}{1-e},
\end{align}
where  $\rmin$ and $\rmax$ are the periapsis and apapsis distance. This geometric choice is convenient since explicit analytic expressions for the orbit and other parameters such as $\nE$ and $\nL$ are known in terms of $p$ and $e$ \cite{Schmidt:2002qk,Fujita:2009bp}. We further adopt the convention that at $q_r=0$ the body is at the apapsis $\rmax$ and $t=\phi=0$.

We will further regularly refer to the orbital frequencies of the orbit as viewed by a distant inertial observer,
\begin{align}
	\Omega_r &= \frac{\Upsilon_r}{\avg{\d{t}{\mt}}},\\
	\Omega_\phi &= \frac{\Upsilon_\phi}{\avg{\d{t}{\mt}}}.
\end{align}
Their main relevance for our present purpose is that the spectrum of gravitational perturbations produced by a particle in an eccentric equatorial orbit is given by all possible integer combinations of $\Omega_r$ and $\Omega_\phi$.

\section{Method}\label{sec:method}
To calculate the self-force on eccentric equatorial orbits in Kerr spacetime, roughly the same methodology will be used as in \cite{vandeMeent:2015lxa} to calculate the regular metric perturbation and redshift invariant. This built on the pioneering work of Keidl, Shah, Friedman et al. \cite{Keidl:2006wk,Keidl:2010pm,Shah:2010bi,Shah:2012gu} culminating in calculations of the red-shift on circular equatorial orbits in Kerr.

The key idea is to avoid the non-separability of the linearized Einstein equation, by solving the separable Teukolsky equation \cite{Teukolsky:1972my,Teukolsky:1973ha} for the Weyl scalar $\psi_4$ instead. In vacuum regions away from the particle orbit, the formalism of Chrzanowski, Cohen, and Kegeles (CCK) \cite{Cohen:1974cm,Chrzanowski:1975wv,Kegeles:1979an} allows given $\psi_4$ the construction of a metric perturbation in the ORG which produces the same $\psi_4$. A key result of Wald \cite{Wald:1973} shows that any two metric perturbations producing the same $\psi_4$ differ by at most a gauge transformation and a perturbation within the four-dimensional Plebanski-Demianski family \cite{Plebanski:1976gy} of vacuum type D metrics.
The gauge independent part of this missing piece of the metric may be recovered analytically by imposing continuity of gauge invariant fields across the particle's orbit \cite{completion,MerlinThesis} (more details follow in Sec.~\ref{sec:completion}).

In this construction, we make sure to only work in the vacuum regions away from the particle source. Since the source term in the Teukolksy equation for individual modes is smeared out over the region between periapsis $\rmin$ and apapsis $\rmax$, this requires us to do the mode-by-mode calculations either in the vacuum region outside $\rmax$ or inside $\rmin$. Only at the last stage before summing the modes to obtain the self-force are these results analytically extended to the particle location. There are a number of reasons for applying this method of ``extended homogeneous solutions''.
\begin{itemize}
\item The CCK procedure is only well-defined for vacuum perturbations of the background. Hence it cannot be applied mode-by-mode on full solutions of the Teukolsky equation in the libration region where it does not have a vacuum source.
\item By doing the reconstruction in the inside and outside vacuum regions separately, we automatically enforce regularity at the horizon and infinity respectively. Consequently, analytic extension will automatically produce the right one-sided limits towards the particle to be used in the averaged mode-sum formula \eqref{eq:modesumavg}.
\item Finally, using the extended homogeneous modes avoids the Gibbs phenomenon that prevents uniform convergence of the sum over Fourier modes of the metric field in a neighbourhood of the particle \cite{Barack:2008ms}. 
\end{itemize}

In the remainder of this paper solutions in the outside vacuum region (or analytic extensions thereof) are labelled with $\pI$. Similarly, solutions in the vacuum region inside the particle (or analytic extensions thereof) are labelled with $\pH$. In the following subsections we will review each of the key steps in this procedure, with particular focus on the aspects of the method that differ from \cite{vandeMeent:2015lxa}.

\subsection{Weyl scalar \texorpdfstring{$\psi_4$}{psi4}}
The first step in our calculation is to determine the linear perturbation to the Weyl scalar,
\begin{align}
\psi_4 &=C_{\alpha\beta\gamma\delta}\tet{2}{\alpha}\tet{4}{\beta}\tet{2}{\gamma}\tet{4}{\delta} = C_{2424},\\
 &=\psi_4^{(0)} + \mr\psi_4^{(1)} +\bigO(\mr^2).
\end{align}
In Kerr spacetime $\psi_4^{(0)}=0$, and with some abuse of notation we will drop the superscript and refer to the (normalized) linear perturbation $\psi_4^{(1)}$ as simply $\psi_4$. Teukolsky's classical result \cite{Teukolsky:1972my,Teukolsky:1973ha} is that the equations of motion for $\psi_4$ in algebraicly special spacetimes (such as Kerr) decouple from the other components of the curvature. Moreover, the resulting equation can be solved by separation of variables. In the inside and outside vacuum regions, the solution to the Teukolsky equation can be written,
\begin{equation}\label{eq:psi4exp}
\psi_{4}^\pm = \frac{\rho^{4}}{\sqrt{2\pi}}\sum_{\spl m\omega} Z_{\spl m\omega}^\pm \R[\pm]{-2}{\spl m\omega}(r) \SWSH{-2}{\spl m\omega}(z)\ee^{\ii m\phi-\ii\omega t},
\end{equation}
where $\rho$ is one of the Newman-Penrose spin coefficients (see Appendix \ref{app:A}), and the sum over $\omega$ is over all possible integer combinations $m\Omega_\phi+ n\Omega_r$ of the azimuthal ($\Omega_\phi$) and radial ($\Omega_r$) orbital frequencies (with respect to Boyer-Lindquist coordinate time). Furthermore, in the above expansion the $\SWSH{s}{\spl m\omega}(z)$ are spin-weighted spheroidal harmonics of spin-weight $s$ satisfying the angular equation,
\begin{align}
\begin{split}
\hh{
	\d{}{z}\Bh{(1-z^2)\d{}{z}}
-	U_{s\spl m\omega}(z)
}\SWSH{s}{\spl m\omega}(z) = 0,
\end{split}\label{eq:SWSHeq}
\end{align}
with the potential
\begin{align}
U_{s\spl m\omega} = \frac{(m+sz)^2}{1-z^2} -(a\omega z-s)^2 +s(s-1)-\!\SEV{s}{\spl m\omega},
\end{align}
where $\SEV{s}{\spl m\omega}$ is the angular separation constant; the $\R[\pm]{s}{\spl m\omega}(r)$ are solutions of the homogeneous radial Teukolsky equation,
\begin{align}
&
\hh{
	\Delta^{-s}\d{}{r}\Bh{\Delta^{s+1}\d{}{r}}
	-V_{s\spl m\omega}(r)
}\R{s}{\spl m\omega}(r) 
= 0,
\label{eq:radteukeq}
\end{align}
with potential
\begin{align}
&V_{s\spl m\omega} =\TEV{s}{\spl m\omega}-4\ii s\omega r -\frac{K_{m\omega}^2-2\ii s (r-1)K_{m\omega}}{\Delta},
\end{align}
where
\begin{align}
 K_{m\omega} 			&\equiv (r^2+a^2)\omega- a m,\\
 \TEV{s}{\spl m\omega} &\equiv \SEV{s}{\spl m\omega} + a^2\omega^2-2ma\omega,
\end{align}
which satisfy physical retarded boundary conditions at infinity ($\pI$) or at the horizon ($\pH$). Finally, the coefficients $Z_{\spl m\omega}^\pm$ can be determined using variation of parameters,
\begin{equation}\label{eq:sourceint}
Z_{\spl m\omega}^\pm = \int_{r_{\mathrm{min}}}^{r_{\mathrm{max}}}  \frac{
	\R[\mp]{-2}{\spl m\omega}(r)\T{-2}{\spl m\omega}(r)
	}{
	W[\R[\pI]{-2}{lm\omega},\R[\pH]{-2}{\spl m\omega}](r)
	}
	\id{r},
\end{equation}
where $W[R_1,R_2]$ is the Wronskian of two homogeneous solutions and $\T{-2}{\spl m\omega}(r)$ is the source for the radial Teukolsky equation for a point particle of a geodesic, for which explicit expressions confirming to our sign conventions can be found in the forthcoming \cite{Meent:2015a}.

\subsection{Hertz potential}
To reconstruct the metric perturbation, we first need to construct an intermediate quantity known as the Hertz potential. In the outgoing radiation gauge, the Hertz potential satisfies a fourth-order differential equation with $\psi_4$ appearing as a source term,
\begin{align}\label{eq:psiORGrad}
\rho^{-4}\psi_4 &=\frac{1}{32}\Delta^2 (\mathcal{D}^{\dagger})^4 \Delta^2 \bar{\Psi}_\textrm{ORG},
\end{align}
where
\begin{align}
\mathcal{D}^{\dagger} &= \partial_r- \frac{(r^2+a^2)\partial_t+a\partial_\phi}{\Delta}.
\end{align}
A key feature of Eq. \eqref{eq:psiORGrad} is that it features no $z$ derivatives, leading to it being called the \emph{radial equation} for $\Psi_\textrm{ORG}$. There also exists an angular equation $\Psi_\textrm{ORG}$ linking it to $\psi_0$ \cite{Lousto:2002em}. That equation was used in \cite{Shah:2012gu} to calculate self-force corrections to circular orbits in Kerr.

In vacuum regions, the ORG Hertz potential must also satisfy the homogeneous Teukolsky equation for $s=+2$ fields. Consequently, in the interior and exterior vacuum regions  $\Psi_{ORG}$ can be decomposed in spin-weighted spheroidal harmonic frequency modes,
\begin{equation}\label{eq:PsiExp}
\Psi_{ORG}^{\pm} = \frac{1}{\sqrt{2\pi}}\sum_{\spl m\omega} \Psi_{\spl m\omega}^\pm \R[\pm]{2}{\spl m\omega}(r)\; \SWSH{2}{\spl m\omega}(z)\ee^{\ii m\phi-\ii\omega t}.
\end{equation}
Consequently, both sides of Eq. \eqref{eq:psiORGrad} can be expanded in modes. As observed by Ori \cite{Ori:2002uv} for the radial equation linking $\psi_0$ and $\Psi_\textrm{IRG}$, this equation decouples into individual equations for all the modes, which can easily be inverted by looking at the asymptotic limit towards infinity and the horizon. The inversion was solved explicitly in \cite{vandeMeent:2015lxa} for the case at hand, yielding an algebraic relation between the coefficients $\Psi_{\spl m\omega}^\pIH$ and  $Z_{\spl m\omega}^\pIH$,
\begin{widetext}
\begin{align}\label{eq:psiI}
(-1)^{\spl+m}\frac{\Psi_{\spl m\omega}^\pI}{Z_{\spl m\omega}^\pI} &=\begin{cases}
\frac{2}{\omega^{4}} 
\hspace{181pt}\phantom{.}&\text{for $\omega\neq 0$,}
\\
 \frac{32}{\rule{0pt}{7pt}
(\frac{ma}{\kappa}-2\ii)(\frac{ma}{\kappa}-\ii)(\frac{ma}{\kappa})(\frac{ma}{\kappa}+\ii)
}
&\text{for $\omega= 0$ but $ma\neq 0$,}
\\
- 32  
&\text{for $\omega=ma= 0$,}
\end{cases}
\\
\label{eq:psiH}
(-1)^{\spl+m}\frac{\Psi_{\spl m\omega}^\pH}{Z_{\spl m\omega}^\pH } &=\begin{cases}
 512\kappa^4
\frac{
	(\frac{ma}{\kappa}-2\omega-2\ii)(\frac{ma}{\kappa}-2\omega-\ii)(\frac{ma}{\kappa}-2\omega)(\frac{ma}{\kappa}-2\omega+\ii)
}{
	 p_{\spl m\omega}
}
&\text{for $\omega\neq 0$,}
\\[3pt]
32\frac{
	(\frac{ma}{\kappa}-2\ii)(\frac{ma}{\kappa}-\ii)(\frac{ma}{\kappa})(\frac{ma}{\kappa}+\ii)
}{\rule{0pt}{10.5pt}
	\kappa^4 \bh{(\spl-1)\spl(\spl+1)(\spl+2)}^2
} 
&\text{for $\omega= 0$ but $ma\neq 0$,}
\\
  \frac{32}{\rule{0pt}{10.5pt}\bh{(\spl-1)\spl(\spl+1)(\spl+2)}^2} 
&\text{for $\omega=ma= 0$,}
\end{cases}
\end{align}
\end{widetext}
where
\begin{equation}\label{eq:TSconstant}
\begin{split}
p_{\spl m\omega} = &\bh{(\TEV{-2}{\spl m\omega} + 2)^2 + 4 m a \omega - 
     4 a^2 \omega^2}
     \\
     &\times\bh{\TEV{-2}{\spl m\omega}^2 + 36 m a \omega - 
     36 a^2  \omega^2}
     \\
     &+ (2 \TEV{-2}{\spl m\omega} + 3) (96 a^2 \omega^2 - 
     48 m a \omega)\\
     &+ 144 \omega^2 (1 - a^2)
\end{split}
\end{equation}
is the Teukolsky-Starobinsky constant, and $\kappa = \sqrt{1-a^2}$.

The expansion \eqref{eq:PsiExp} is somewhat impractical to work with because there are no analytically known spin raising and lowering operators for spheroidal harmonics, making analytical manipulation of its derivatives (as will be required shortly) impossible. It is therefore useful to expand the spin-weighted spheroidal harmonics in spin-weighted spherical harmonics using
\begin{equation}
\SWSH{s}{\spl m\omega}(z)=\sum_{l}\Sb{s}{m\omega}{\spl}{l} \Y{s}{lm}(z),
\end{equation}
where the mixing coefficients $\Sb{s}{m\omega}{\spl}{l}$ can be calculated numerically \cite{Hughes:1999bq}, and are known to decay exponentially with $\abs{\spl-l}$.

The resulting expansion in spin-weighted spherical harmonics,
\begin{equation}\label{eq:PsiY}
\begin{split}
\Psi_{ORG}^\pm = \frac{1}{\sqrt{2\pi}}\sum_{l\spl m\omega} \Psi_{\spl m\omega}^\pm \R[\pm]{2}{\spl m\omega}(r)\; \Sb{2}{m\omega}{\spl}{l}\\
\times \Y{2}{lm}(z)\ee^{\ii m\phi-\ii\omega t},
\end{split}
\end{equation}
will be the starting point of the following subsections.

\subsection{Reconstructed metric}
We now turn to reconstructing the metric perturbation. The non-vanishing tetrad components of the metric perturbation are given by
\begin{alignat}{3}
h_{11} &\equiv \tet{1}{\mu}\tet{1}{\nu}h_{\mu\nu} &&= \MROP{11}^{ORG}\Psi_{ORG}+ c.c.,
\\
h_{13} &\equiv \tet{1}{\mu}\tet{3}{\nu}h_{\mu\nu} &&= \MROP{13}^{ORG}\Psi_{ORG},
\\
h_{33} &\equiv \tet{3}{\mu}\tet{3}{\nu}h_{\mu\nu} &&= \MROP{33}^{ORG}\Psi_{ORG},
\\
h_{14} &\equiv \tet{1}{\mu}\tet{4}{\nu}h_{\mu\nu} &&=  \bar{h}_{13},\text{ and}
\\
h_{44} &\equiv \tet{4}{\mu}\tet{4}{\nu}h_{\mu\nu} &&=\bar{h}_{33},
\end{alignat}
with\footnote{The astute reader will notice that the expressions below differ from Eqs. (103)-(105) in \cite{vandeMeent:2015lxa} by an overall minus sign. This change is due to a different sign convention for $\psi_4$ used there.}
\begin{align}
\MROP{11}^{ORG} &= -\rho^{-4}\bh{\bar\delta-3\alpha-\bar\beta+5\varpi}\bh{\bar\delta-4\alpha+\varpi},
\displaybreak[0]\\
\MROP{13}^{ORG} &= -\frac{\rho^{-4}}{2}\cbB{
	\bh{\bar\delta-3\alpha+\bar\beta+5\varpi+\bar\tau}\bh{\hat\Delta+\mu-4\gamma}
	\\
	&\quad+
	\bh{\hat\Delta+5\mu-\bar\mu-3\gamma-\bar\gamma}\bh{\bar\delta-4\alpha+\varpi}
	 },\text{ and}\nonumber
\\
\MROP{33}^{ORG} &= -\rho^{-4}\bh{\hat\Delta+5\mu-3\gamma+\bar\gamma}\bh{\hat\Delta+\mu-4\gamma},
\end{align}
where ``+c.c.'' represents the complex conjugate of the preceding terms, and $\bar\delta = \tet{4}{\mu}\partial_\mu$, $\hat\Delta = \tet{2}{\mu}\partial_\mu$, and the remaining Greek symbols represent the Newman-Penrose spin-coefficients. Their values are given explicitly in Appendix \ref{app:A}.

The coordinate components of $h$ are reconstructed as
\begin{equation}
h_{\mu\nu} = \itet{a}{\mu}\itet{b}{\nu}h_{ab}.
\end{equation}
Applying the above formula mode-by-mode to the expansion \eqref{eq:PsiExp}, substituting all $z$ derivatives by spin-lowering operators,
\begin{equation}
\SWL{s}=\sqrt{1-z^2}\hh{\partial_z+\frac{\ii}{1-z^2}\partial_\phi-\frac{s z}{1-z^2}},
\end{equation}
and using
\begin{equation}\label{eq:SLnorm}
\SWL{s}\Y{s}{lm}(z) = -\sqrt{(l+s)(l-s+1)} \Y{(s-1)}{lm}(z),
\end{equation}
we obtain a mode expansion for the metric perturbation in the interior and exterior vacuum regions,
\begin{equation}\label{eq:hmunu}
\begin{split}
h_{\mu\nu}^\pIH = \sum_{\substack{m{\omega}si\\l\spl}} \Psi_{\spl m\omega}^\pIH\R[\pIH,(i)]{2}{\spl m\omega}(r)\Sb{2}{m\omega}{\spl}{l}\MC_{\mu\nu}^{l m{\omega}si}(r,z)&
\\
\times \Y{s}{lm}(z){\ee}^{\ii m\phi-\ii\omega t}+c.c&.,
\end{split}
\end{equation}
where the $\MC_{\mu\nu}^{l m{\omega}si}(r,z)$ are coefficients that still depend on $r$ and $z$, and whose explicit analytic form is known, but not very illustrative and will not be given here.

\subsection{Gravitational Self-force}\label{sec:GSF}
The expansion for the metric perturbation \eqref{eq:hmunu} can be inserted into \eqref{eq:GSFdef} to obtain the (extended) self-force in the interior and exterior vacuum regions. By analytically extending the homogeneous solutions of the radial Teukolsky equation $\R[\pIH,(i)]{2}{\spl m\omega}$ these expressions can be extended towards the particle worldline. Formally, we write
\begin{align}
\Fext_{\Rad}^{\mu,\pIH} &=P^{\mu\alpha\beta\gamma}\CD{\alpha} h_{\beta\gamma}^\pIH\\
&=\begin{aligned}[t]
 \sum_{\substack{m{\omega}si\\l\spl}} 
\Psi_{\spl m\omega}^\pIH
\R[\pIH,(i)]{2}{\spl m\omega}(r)
\Sb{2}{m\omega}{\spl}{l}
\MC^\mu_{l m{\omega}si}(r,z)&
\\
\times
\Y{s}{lm}(z){\ee}^{\ii m\phi-\ii\omega t}+c.c&.,\label{eq:GSF1}
\end{aligned}
\end{align}
where again we have replaced any $z$ derivatives with spin-lowering operators and the $\MC^\mu_{lm{\omega}si}(r,z)$ are a (new)\footnote{Throughout this discussion we will use the symbol $\MC$ for the coefficients in the various mode expansion, despite its value changing at each step.}  set of analytically known coefficients.

Our next task is to decompose \eqref{eq:GSF1} into $l$-modes, so that we can use it as input for the averaged mode-sum formula \eqref{eq:modesumavg}. Equation \eqref{eq:GSF1} already hints at a mode decomposition, but at this stage we have three problems: it is decomposed in the wrong harmonics for use in Eq. \eqref{eq:modesumavg}, it has not yet incorporated the complex conjugate ($+c.c.$) terms, and the coefficients $\MC$ are still functions of $z$. We will start by remedying the first issue.

Factoring out appropriate factors of $\sqrt{1-z^2}$ from the $\MC$'s and recognizing that $\Y{s}{\abs{s}0}(z) \propto \sqrt{1-z^2}^{\abs{s}}$, we can use the integral product of spin-weighted spherical harmonics (using Wigner 3j-symbols),
\begin{equation}
\begin{split}
\int_{-1}^{1}\Y{s_1}{l_1m_1}(z)&\Y{s_2}{l_2m_2}(z)\Y{s_3}{l_3m_3}(z)\id{z} =\\
 &\sqrt{(2l_1+1)(2l_2+1)(2l_3+1)}\\
&\quad\times\begin{pmatrix}
l_1 & l_2 & l_3\\
m_1 & m_2 & m_3
\end{pmatrix}
\begin{pmatrix}
l_1 & l_2 & l_3\\
s_1 & s_2 & s_3
\end{pmatrix},
\end{split}
\end{equation}
to write
\begin{align}
\Y{2}{l_1m}(z)&=\sum_{l_2}
\frac{
\YA{2}{m}{l_1}{l_2} \Y{}{l_2m}(z)
}{
1-z^2
}
,\label{eq:YA1}\\
\Y{1}{l_1m}(z)&=\sum_{l_2}
\frac{
\YA{1}{m}{l_1}{l_2}\Y{}{l_2m}(z)
}{
\sqrt{(l_1-1)(l_1+2)}\sqrt{1-z^2}
}
,\\
\Y{0}{l_1m}(z)&=\sum_{l_2}\frac{
\YA{0}{m}{l_1}{l_2}\Y{}{l_2m}(z)
}{
\sqrt{(l_1-1)l_1(l_1+1)(l_1+2)}
} ,\\
\Y{-1}{l_1m}(z)&=\sum_{l_2}\frac{
\YA{-1}{m}{l_1}{l_2}
\Y{}{l_2m}(z)
\sqrt{(l_1-2)!}
}{
\sqrt{(l_1+2)!l_1(l_1+1)}\sqrt{1-z^2}
} ,\label{eq:YA4}
\end{align}
with
\begin{align}
&\begin{aligned}
\YA{2}{m}{l_1}{l_2} = 
(-1)^m\sqrt{\frac{8}{3}(2l_1+1)(2l_2+1)}&
\\
\times\begin{pmatrix}
2 & l_1 & l_2\\
0 & m	& -m
\end{pmatrix}
&\begin{pmatrix}
2 & l_1 & l_2\\
-2 & 2	& 0
\end{pmatrix}
,
\end{aligned}
\\
&\begin{aligned}
\YA{1}{m}{l_1}{l_2} = 
&(-1)^{m+1}
\sqrt{2(l_1-1)(l_1+2)(2l_1+1)}
\\
&\quad\times\sqrt{2l_2+1}
\begin{pmatrix}
1 & l_1 & l_2\\
0 & m	& -m
\end{pmatrix}
\begin{pmatrix}
1 & l_1 & l_2\\
-1 & 1	& 0
\end{pmatrix}
,
\end{aligned}\\
&\YA{0}{m}{l_1}{l_2} = \sqrt{\frac{(l_1+2)!}{(l_1-2)!}}\delta_{l_1l_2},\\
\intertext{and}
&\begin{aligned}
\YA{-1}{m}{l_1}{l_2} = 
&(-1)^{m}
\sqrt{2 l_1 (l_1+1)(2l_1+1)\frac{(l_1+2)!}{(l_1-2)!}}
\\
&\;\times\sqrt{2l_2+1}
\begin{pmatrix}
1 & l_1 & l_2\\
0 & m	& -m
\end{pmatrix}
\begin{pmatrix}
1 & l_1 & l_2\\
1 & -1	& 0
\end{pmatrix}
,
\end{aligned}
\end{align}
where the $l_1$ dependent factors in Eqs. \eqref{eq:YA1}-\eqref{eq:YA4} have been introduced to absorb the $l$ dependence of the $\MC$'s (which originated completely from repeated application of \eqref{eq:SLnorm})  in the $\YA{s}{m}{l_1}{l_2}$. The result is an expansion featuring only ordinary spherical harmonics,
\begin{align}\label{eq:GSF2}
\Fext_{\Rad}^{\mu,\pIH} &=\begin{aligned}[t]
  \sum_{\substack{m{\omega}si\\l_1l_2\spl}} 
\Psi_{\spl m\omega}^\pIH
\R[\pIH,(i)]{2}{\spl m\omega}(r)\Sb{2}{m\omega}{\spl}{l_1}
\YA{s}{m}{l_1}{l_2}
\hspace{3em}&
\\
\times
\MC^\mu_{m{\omega}si}(r,z)
\Y{}{l_2m}(z){\ee}^{\ii m\phi-\ii\omega t}+c.c.,&
\end{aligned}
\end{align}
where the definition of the $\MC$'s has again been changed.

To resolve the complex conjugate terms we observe that the individual factors in \eqref{eq:GSF2} have the following symmetry properties under simultaneous complex conjugation and relabelling $(m,\omega)\to (-m,-\omega)$,
\begin{align}
\bar\MC^\mu_{(-m)(-\omega) si}(r,z) &= (-1)^{s+\delta_{z\mu}}\MC_{m\omega si}(r,-z),\\
\bar\Psi_{\spl(-m)(-\omega)}^\pIH&= (-1)^\spl\Psi_{\spl m\omega}^\pIH,\\
\Rb[\pIH,(i)]{2}{l_1(-m)(-\omega)}(r) &= \R[\pIH,(i)]{2}{\spl m\omega}(r),\displaybreak[0]\\
\Sb{2}{(-m)(-\omega)}{\spl}{l}&=(-1)^{\spl+l}\Sb{2}{m\omega}{\spl}{l},\\
\YA{s}{-m}{l_1}{l_2} &= (-1)^{s+l_1+l_2}\YA{s}{m}{l_1}{l_2},\\
\Y{}{l_2(-m)}(z) &= (-1)^m\Y{}{l_2m}(z).
\end{align}
Applying these identities to the complex conjugate terms in \eqref{eq:GSF2}, we obtain
\begin{align}\label{eq:GSF3}
\begin{split}
&\Fext_{\Rad}^{\mu,\pIH} =
\!\!\sum_{\substack{m{\omega}si\\l_1l_2\spl}}\!\!
\Psi_{\spl m\omega}^\pIH
\R[\pIH,(i)]{2}{\spl m\omega}(r)
\Sb{2}{m\omega}{\spl}{l_1} 
\YA{s}{m}{l_1}{l_2}
{\ee}^{\ii m\phi-\ii\omega t}\\
&\hspace{-0.87em}\times
\hh{\MC^\mu_{m{\omega}si}(r,z)+(\umin 1)^{l_2+m+\delta_{z\mu}}\MC^\mu_{m{\omega}si}(r,-z)}
\!\!\Y{}{l_2m}(z).
\end{split}
\end{align}
From the functional dependence of \eqref{eq:GSF3} on $z$ we immediately observe some important symmetry properties of $\Fext_{\Rad}^{\mu,\pIH}$ under reflection in the equatorial plane $z_0=0$. First we observe that $\Fext_{\Rad}^{z,\pIH}$ is an odd function of $z$ and thus vanishes identically on the equator $z_0=0$ as expected. The remaining three components of $\Fext_{\Rad}^{\mu,\pIH}$ are all even.

We now turn our attention to the remaining $z$ dependence of the coefficient functions $\MC$ for the $\mu\neq z$ components. We start by taking the limit towards $t=t_0$ and $r=r_0$, and taking the average of the inside $\pH$ and outside $\pI$ values, where it is understood that the $\pH$ limit is taken from the inside and vice versa for the $\pI$ limit. We obtain 
\begin{equation}
\begin{split}\label{eq:GSF4}
\Fext_{\Rad}^{\mu,\Avg} =
 \sum_{\substack{\omega si\\l+m\\ \mathrm{even}}} 
\chi^{\Avg}_{lm\omega s i}\MC^\mu_{m{\omega}si}(z^2)\Y{}{lm}(z,\phi)
\quad\quad&\\
+z\sum_{\substack{\omega si\\l+m\\ \mathrm{odd}}} 
\chi^{\Avg}_{lm\omega s i}\tilde\MC^\mu_{m{\omega}si}(z^2)\Y{}{lm}(z,\phi)
,&
\end{split}
\end{equation}
where the even/odd structure has been made explicit,
\begin{equation}\label{eq:chisymb}
\begin{split}
\chi^{\Avg}_{lm\omega s i} =
\sum_{l_1\spl} 
\tfrac{
	\Psi_{\spl m\omega}^\pI
	\R[\pI,(i)]{2}{\spl m\omega}(r_0)
	+
	\Psi_{\spl m\omega}^\pH
	\R[\pH,(i)]{2}{\spl m\omega}(r_0)
}{
2
}&\\
\times\Sb{2}{m\omega}{\spl}{l_1} 
\YA{s}{m}{l_1}{l}
{\ee}^{-\ii\omega t_0},&
\end{split}
\end{equation}
and $\MC$ and $\tilde{\MC}$ are smooth functions of $z^2$. Note that due to the singular nature of $\Fext_{\Rad}^{\mu,\Avg}$ near the particle, we do not necessarily expect the series in \eqref{eq:GSF4} to converge pointwise. Nonetheless, they are still expected to converge in a generalized (distributional) sense to a unique function that is smooth everywhere in a neighbourhood of $(z_0,\phi_0)$ except at $(z_0,\phi_0)$ itself.

Since $\MC$ and $\tilde{\MC}$ are smooth functions we can expand them in a Taylor series in $z^2$,
\begin{align}
\MC^\mu_{m{\omega}si}(z^2) &=\sum_{k=0}^\infty \MC^{\mu,k}_{m{\omega}si} z^{2k},\\
\tilde\MC^\mu_{m{\omega}si}(z^2) &=\sum_{k=0}^\infty \tilde\MC^{\mu,k}_{m{\omega}si} z^{2k},
\end{align}
yielding
\begin{equation}
\begin{split}\label{eq:GSF5}
\Fext_{\Rad}^{\mu,\Avg} =
 \sum_{k=0}^\infty\Bh{z^{2k}\sum_{\substack{\omega si\\l+m\\ \mathrm{even}}} 
\chi^{\Avg}_{lm\omega s i}\MC^{\mu,k}_{m{\omega}si}\Y{}{lm}(z,\phi)
\quad\quad&\\
+z^{2k+1}\sum_{\substack{\omega si\\l+m\\ \mathrm{odd}}} 
\chi^{\Avg}_{lm\omega s i}\tilde\MC^{\mu,k}_{m{\omega}si}\Y{}{lm}(z,\phi)
}.&
\end{split}
\end{equation}
A key observation at this point is that the coefficients $\MC^{\mu,k}_{m{\omega}si}$ and $\tilde\MC^{\mu,k}_{m{\omega}si}$ are independent of $l$. Consequently, the singular behaviour near $(z_0,\phi_0)$ of the sums over $l$ and $m$ in \eqref{eq:GSF5} is limited by the large $l$ behaviour of $\chi^{\Avg}_{lm\omega s i}\Y{}{lm}(z,\phi)$. The analysis of \cite{Pound:2013faa} implies that this combination has to decay by at least $l^{-1}$ as $l\to\infty$. Consequently, the singular structures of the sums in \eqref{eq:GSF5} are at worst,
\begin{align}
\sum_{\substack{\omega si\\l+m\\ \mathrm{even}}}
\chi^{\Avg}_{lm\omega s i}\MC^{\mu,k}_{m{\omega}si}\Y{}{lm}(z,\phi_0) \propto \delta(z)+ \bigO(\log\abs{z}),
\end{align}
and
\begin{align}\label{eq:oddsing}
\sum_{\substack{\omega si\\l+m\\ \mathrm{odd}}}
\chi^{\Avg}_{lm\omega s i}\tilde\MC^{\mu,k}_{m{\omega}si}\Y{}{lm}(z,\phi_0) \propto \bigO(z^{-1}).
\end{align}
This implies that the $k\geq 1$ terms in \eqref{eq:GSF5} are $\bigO(z)$. However, the mode-sum formula is insensitive to contributions to $\Fext^{\mu}$ of order $\bigO(z)$. Consequently, we can drop the $k \geq 1$ terms to obtain,
\begin{equation}
\begin{split}\label{eq:GSF6}
\Fext_{\Rad}^{\mu,\Avg} =
&\sum_{\substack{\omega si\\ l+m\\ \mathrm{even}}} \!\!
\chi^{\Avg}_{lm\omega s i}\MC^{\mu,0}_{m{\omega}si}\Y{}{lm}(z,\phi)
\\
&+
\sum_{\substack{\omega si\\ l+m\\ \mathrm{odd}}} \!\!
\chi^{\Avg}_{lm\omega s i}\tilde\MC^{\mu,0}_{m{\omega}si}z\Y{}{lm}(z,\phi)
 + \bigO(z).
\end{split}
\end{equation}
Dropping the $k \geq 1$ terms amounts to choosing an alternative extension for $\Fext_{\Rad}^{\mu,\Avg}$ that is compatible with the extension used for calculating the Lorenz gauge regularization parameters. The definition of this extension, however, is deeply entwined with the specifics of the metric reconstruction procedure. Hence it does not have a straightforward characterization at the level of the singular field. This provides a further\footnote{Lack of knowledge of the gauge transformation linking radiation gauge and Lorenz gauge solutions beyond leading order already provides a significant obstacle for such an undertaking.} roadblock to analytically calculating ``higher-order regularization parameters'' compatible with this method as was done for Lorenz gauge methods in \cite{Heffernan:2012su} and \cite{Heffernan:2012vj}.

In practice we actually observe that the sum over $l+m$ odd modes is less singular than the worst case scenario indicated in \eqref{eq:oddsing}, and the $k=0$ term only produces a $\bigO(z)$ contribution to the (extended) self-force. If we were to drop that term from \eqref{eq:GSF6}, the remaining term has the form of an explicit expansion in $\Y{}{lm}$ modes, and we would be able to read-off the $l$-modes directly. However, we have thus far not been able to prove this empirical observation analytically. We therefore proceed by observing that $z\propto \Y{}{10}(z,\phi)$ to expand the product $z\Y{}{lm}(z,\phi)$ using
\begin{equation}
z\Y{}{l_1m}(z,\phi)=\sum_{l_2}\YB{}{m}{l_1}{l_2}
 \Y{}{l_2m}(z,\phi),
\end{equation}
with
\begin{equation}
\begin{split}
\YB{}{m}{l_1}{l_2} = (-1)^{m+l_1+1}&(l_1-l_2)
\\
\times&\sqrt{\frac{l_1+l_2+1}{2}}
\begin{pmatrix}
1 & l_1 & l_2\\
0 & m	& -m
\end{pmatrix}.
\end{split}
\end{equation}
The result is an expansion of $\Fext_{\Rad}^{\mu,\Avg}$ in spherical harmonics with coefficients that do not depend on $z$,
\begin{equation}
\begin{split}\label{eq:GSF7}
\Fext_{\Rad}^{\mu,\Avg} =
&\sum_{\substack{\omega si\\ lm}}
\Bh{
\chi^{\Avg}_{lm\omega s i}\MC^{\mu,0}_{m{\omega}si}
\\
&\hspace{2em}+\sum_{l_2} \chi^{\Avg}_{l_2m\omega s i}\YB{}{m}{l_2}{l}\tilde\MC^{\mu,0}_{m{\omega}si}
}
\Y{}{lm}(z,\phi).
\end{split}
\end{equation}
Consequently, the $l$-modes needed as input for the averaged mode-sum formula \eqref{eq:modesumavg} can  be directly read-off,
\begin{equation}
\begin{split}\label{eq:GSFlmodes}
F_{\Rad,l}^{\mu,\Avg} =
&\sum_{\substack{m\omega si}} \Bh{
\chi^{\Avg}_{lm\omega s i}\MC^{\mu,0}_{m{\omega}si}
\\
&\hspace{2em}+\sum_{l_2} \chi^{\Avg}_{l_2m\omega s i}\YB{}{m}{l_2}{l}\tilde\MC^{\mu,0}_{m{\omega}si}
}
\Y{}{lm}(0,\phi_0).
\end{split}
\end{equation}
The coefficients $\MC^{\mu,0}_{m{\omega}si}$ and $\tilde\MC^{\mu,0}_{m{\omega}si}$ are analytic functions of the orbital parameters $a$, $\nE$ and $\nL$ and the position along the orbit $q_r$. It is straightforward to obtain their analytical form by explicitly keeping track of the coefficients in the above procedure. Their explicit form is not particularly elucidating and would take a good number of pages to print. We therefore do not give them here, but provide them as a digital supplement to this paper \cite{SD}. We will suffice with noting that it can be explicitly checked that the expression in the supplement satisfy
 \begin{equation}
u_\mu\MC^{\mu,0}_{m{\omega}si}=u_\mu\tilde\MC^{\mu,0}_{m{\omega}si}=0
 \end{equation}
 for all $m$, $\omega$, $s$, and $i$. Consequently, it is automatically ensured that $u_\mu F^\mu=0$, i.e.  the gravitational self-force conserves the rest mass of the particle.

\subsection{Completion}\label{sec:completion}
The final step in our calculation is to determine the contribution to the self-force from the piece of the metric perturbation $h$ that cannot be recovered by the CCK procedure because it is in the kernel of the differential operator that produces $\psi_4$. Wald has shown \cite{Wald:1973} that the gauge invariant content of this kernel is exactly given by perturbations of the Kerr background in the four-dimensional family of Plebanski-Demianski type-D vacuum metrics \cite{Plebanski:1976gy}. Requiring regularity at either infinity or the black hole horizon further reduces this to perturbations within the Kerr family of metrics \cite{Keidl:2010pm}.

For the purpose of calculating gauge invariant quantities, we can thus suffice by completing our metric perturbation reconstructed from $\psi_4$ with a contribution
\begin{equation}
h^{\mathrm{comp},\pIH}_{\mu\nu} = c_M^\pIH h^M_{\mu\nu} + c_J^\pIH h^J_{\mu\nu},
\end{equation}
where if $g_{\mu\nu}(M,J;x)$ represents the Kerr family of metrics as a function of mass $M$ and angular momentum $J$,
\begin{align}
 \label{eq:hM} h^M_{\mu\nu}(x) &\equiv \pd{g_{\mu\nu}(M,J;x)}{M}\Biggr|_{\substack{M=1\\J=a}},\\
 \label{eq:hJ} h^J_{\mu\nu}(x) &\equiv \pd{g_{\mu\nu}(M,J;x)}{J}\Biggr|_{\substack{M=1\\J=a}},
\end{align}
and $c_M^\pIH$ and $c_J^\pIH$ are real numbers.

Determining the completion thus reduces to determining the four numbers,  $c_M^\pIH$ and $c_J^\pIH$. In the outside vacuum region $c_M^\pI$ and $c_J^\pI$ can be determined by fixing the total ADM mass and angular momentum of the system to their physical value giving $c_M^\pI=\nE$ and $c_J^\pI=\nL$. The values of $c_M^\pH$ and $c_J^\pH$ can then be determined by constructing some gauge invariant fields from the total completed metric perturbation that (unlike $\psi_0$ and $\psi_4$) are sensitive to the values of $c_M^\pIH$ and $c_J^\pIH$, and demanding that they are smooth functions of spacetime as one passes from the inside to the outside vacuum regions while avoiding the orbital plane \cite{completion,MerlinThesis}. For equatorial orbits the result of this rigorous calculation coincides with the naive expectation that $c_M^\pH=c_J^\pH=0$.

Note that the completion metrics $h^M_{\mu\nu}$ and $h^J_{\mu\nu}$ are perfectly regular at the particle position. Consequently, the self-force contribution from the completion can be calculated independently from any evaluation of the mode-sum. Applying formula \eqref{eq:GSFdef} directly to Eqs. \eqref{eq:hM} and \eqref{eq:hJ} yields
\begin{equation}
F_{\mathrm{comp}}^{\mu,\Avg} = \frac{\nE}{2} F_{\mathrm{comp}}^{\mu,M} + \frac{\nL}{2} F_{\mathrm{comp}}^{\mu,J}.
\end{equation}
The specific form of the functions $F_{\mathrm{comp}}^{\mu,M}$ and $F_{\mathrm{comp}}^{\mu,J}$ is given in Appendix \ref{app:GSFcomp}.

The above construction reproduces enough of the metric perturbation to calculate any gauge invariant quantities. It should, however, be noted that most of the ``invariant'' quantities studied in the literature on small mass-ratio binaries involve shifts of the orbital frequencies in someway or other. These are actually pseudo-invariants in the sense that they are invariant only under a restricted class of gauge transformations. Unfortunately, the gauge in which we calculate the self-force does not belong to the class of ``suitably smooth and asymptotically flat'' gauges usually considered. More specifically, the ``no-string'' radiation gauge used is discontinuous on a hypersurface separating the black hole horizon from infinity and containing the particle worldline.

In principle, we can find our results in a suitable gauge with a method much similar to the method used for fixing the gauge invariant part of the completion. In this case we need to find some pseudo-invariant fields that are sensitive to the gauge transformations that can change orbital frequencies. We can then find the corrections to the desired gauge by requiring these pseudo-invariant fields to be sufficiently smooth across the particle orbit. This procedure is fairly straightforward on a Schwarzschild background and was discussed in \cite{Shah:2015nva}. The situation on a Kerr background is slightly more involved, mostly due to the fact that completion on Kerr background does not consist purely of $l=0$ and $l=1$ modes. A detailed treatment of the Kerr case will be given in \cite{gaugecompletion}.

\section{Numerical Implementation}\label{sec:implementation}
The numerical implementation of our calculations is practically identical to the implementation used in \cite{vandeMeent:2015lxa} to calculate the redshift invariant. Following \cite{Fujita:2004rb,Fujita:2009us,Throwethesis}, we solve the homogeneous Teukolsky equation for $\psi_4$ using the semi-analytical series solutions of Mano, Suzuki, and Takasugi (MST)\cite{Mano:1996gn,Mano:1996vt}. The full details of our arbitrary precision implementation will appear in a separate paper~\cite{Meent:2015a}. 

The integrals \eqref{eq:sourceint} for the coefficients $Z_{\spl m\omega}^\pm$ of the inhomogeneous solutions are replaced by suitable integrals over the orbital torus described by $q_r$ and $q_z$ as per \cite{Drasco:2003ky,Drasco:2005kz} (In the specific case of equatorial orbits considered here the integrals over $q_z$ are trivial). The integrands for these integrals are smooth functions on this torus. This means that simple trapezoidal numerical integration has spectral convergence, which we exploit following \cite{Fujita:2009us}. An in depth analysis of the spectral convergence of trapezoidal methods for these integrals has recently appeared in \cite{Hopper:2015jxa}.

From the (extended) inhomogeneous solution for $\psi_4$, we obtain the spin-weighted spheroidal modes of the Hertz potential and their radial derivatives evaluated at the particle orbit as a function of $q_r$. Equations.~\eqref{eq:GSFlmodes} and \eqref{eq:chisymb} described (spherical harmonic) $\spl$-modes as a linear combination of these Hertz potential modes using the (infinite dimensional) matrices $\Sb{2}{m\omega}{\spl}{l_1}$, 
$\YA{s}{m}{l_1}{l_2}$ and $\YB{}{m}{l_1}{l_2}$. The matrices $\YA{s}{m}{l_1}{l_2}$ and $\YB{}{m}{l_1}{l_2}$ are evaluated using the explicit expressions in Sec.~\ref{sec:GSF}, whereas $\Sb{2}{m\omega}{\spl}{l_1}$ is evaluated using the method appearing in the appendix of \cite{Hughes:1999bq}. The transformations $\YA{s}{m}{l_1}{l_2}$ and $\YB{}{m}{l_1}{l_2}$ have a finite bandwith and $\Sb{2}{m\omega}{\spl}{l_1}$ decays exponentially with $\abs{\spl-l_1}$ off the diagonal. Consequently, limiting the input to modes of the Hertz potential with an $\spl$ of maximally $\llmax$ introduces an estimable error in GSF $l$-modes that grows exponentially as $l$ approaches $\llmax$. We keep track of this error and discard the $F^\mu_l$ above a maximal value $\lmax$ once the error exceeds a chosen threshold value.

The next numerical task is to evaluate the sum over $\omega$ and $m$ in Eq. \eqref{eq:GSFlmodes}. For this purpose we write \eqref{eq:GSFlmodes} as a nested sum
\begin{align}
F_{l}^{\mu,\Avg} 	&= \sum_{m} F_{lm}^{\mu,\Avg},\\
F_{lm}^{\mu,\Avg}	&=\sum_\omega F_{lm\omega}^{\mu,\Avg},\text{ and}\\
\begin{split}\label{eq:Flmw}
F_{lm\omega}^{\mu,\Avg} &=
\sum_{\substack{si}} \Bh{
\chi^{\Avg}_{lm\omega s i}\MC^{\mu,0}_{m{\omega}si}
\\
&\hspace{2em}+\sum_{l_2} \chi^{\Avg}_{l_2m\omega s i}\YB{}{m}{l_2}{l}\tilde\MC^{\mu,0}_{m{\omega}si}
}
\Y{}{lm}(0,\phi_0).
\end{split}
\end{align}
The conceptually logical thing to do would be to fix $l$ and then calculate all $F_{lm\omega}^{\mu,\Avg}$ contributing to $F_{l}^{\mu,\Avg}$ (until some target threshold is met), and repeat this for the different $l$'s. However, the matrix nature of $\Sb{2}{m\omega}{\spl}{l_1}$, $\YA{s}{m}{l_1}{l_2}$, and $\YB{}{m}{l_1}{l_2}$, means that numerically it is much more efficient to fix $m$ and $\omega$ and then obtain all  $F_{lm\omega}^{\mu,\Avg}$ with that $m$ and $\omega$. Therefore in practice we loop over $m$ and $\omega$ building up all $F_{l}^{\mu,\Avg}$ simultaneously.

We start our loop by picking a value for $\llmax$ and a target precision $\TP$. Consequently, $m$ can only take values between $-\llmax$ and $\llmax$, while for fixed $m$ the frequency $\omega$ takes values in  $\{n \Omega_r + m \Omega| n\in\ZZ\}$. Since the $F_{l}^{\mu,\Avg}$ are real functions it follows that 
\begin{equation}
\bar{F}_{lm\omega}^{\mu,\Avg} = F_{l(-m)(-\omega)}^{\mu,\Avg}.
\end{equation}
Hence we can restrict our attention to modes with $n\geq 0$. For fixed $m$ we start by calculating all $l$ modes with $n=0$ and continue by incrementally increasing $n$ until 
\begin{equation}
\max_{l,\mu} \frac{\norm{F_{lm\omega}^{\mu,\Avg}}_\infty}{\norm{B^\mu}_2} < \TP ,
\end{equation}
where the sup-norm $\norm{\cdot}_\infty$ and the $L_2$ norm $\norm{\cdot}_2$ of the regularization parameter $B$ are taken with respect to the dependence on the orbital phase $q_r$.

After obtaining the $F_{l}^{\mu,\Avg}$ mode, the final step is to subtract the regularization parameters and add the modes together. We use the regularization parameters as given in \cite{Barack:2009ux}, which uses an extension of the full self-force that is compatible with the one used here.\footnote{Note that there is a typo in the expresion for $A^\mu$ in \cite{Barack:2009ux}. A correct expression for $A^\mu$ appears in \cite{Barack:2002mh}, which uses a different extension that does not affect the value of $A^\mu$.} The ``regularized'' $l$-modes
\begin{equation}\label{eq:regF}
F_{l,\reg}^{\mu,\Avg} = F_{l}^{\mu,\Avg} - B^\mu - \frac{C^\mu}{l+1/2},
\end{equation}
should decay with $(l+1/2)^{-2}$. Consequently, the sum of the $l$-modes converges but rather slowly. Convergence can be accelerated by fitting for the large-$l$ behaviour of the series using the known terms. We here use the same procedure as described in \cite{vandeMeent:2015lxa}, where a polynomial in $l^{-1}$ is fit to the partial sums of the series to obtain an estimate for the sum.

\section{Tests and results}\label{sec:results}
\begin{figure}
\includegraphics[width=\columnwidth]{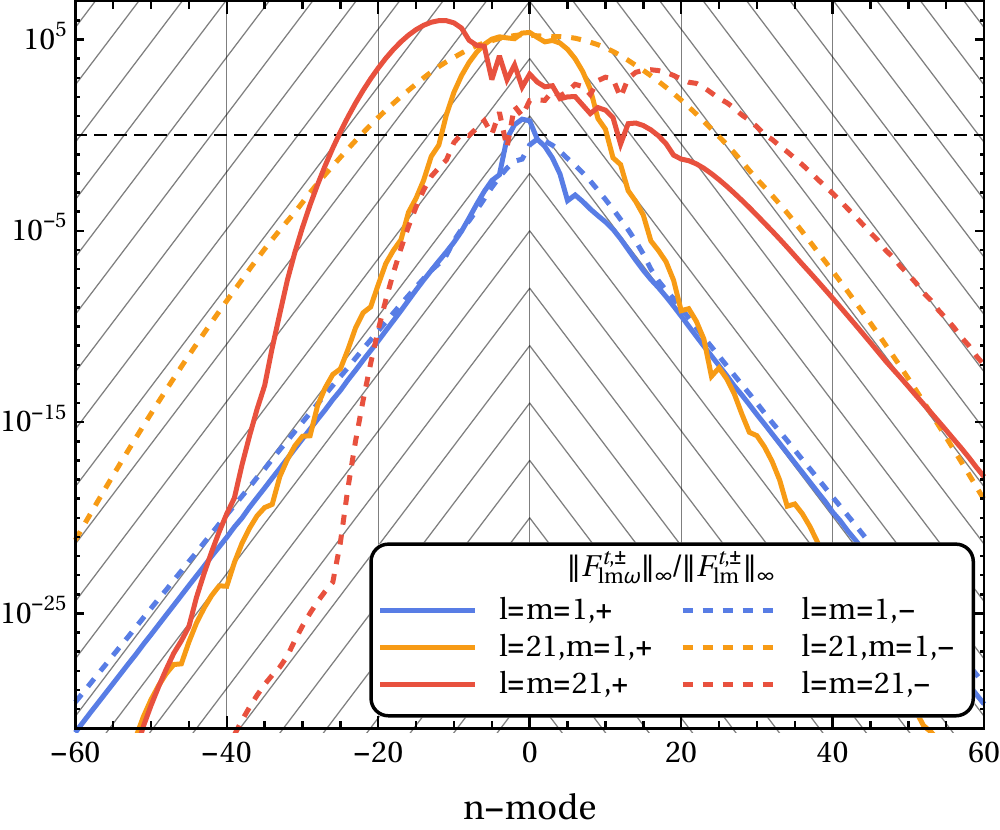}
\caption{Illustration of the convergence of the frequency modes for large radial mode number $\abs{n}$. The plotted lines correspond to the sup-norms of $F^t_{lm\omega}$ obtained for an orbit with $(a,p,e)=(0.9,5.5,0.3)$ and normalized by the the sup-norm of the complete time domain mode $F^t_{lm}$. Solid lines represent values obtained from the outside field, whereas dashed lines represent inside values. The different colors indicate various combinations of $l$ and $m$. The diagonal gridlines indicate a reference decay of $e^{\abs{n}}$.
}\label{fig:nconvplot}
\end{figure}
With all the components for calculating the self-force on eccentric equatorial orbits around a Kerr black hole in place, we can start to do numerical calculations. In this section we present the results of various numerical computations that test the consistency of our methods. We first consider the reconstruction of the time domain $lm$-modes $F_{lm\omega}^{\mu,\Avg}$ from the extended homogeneous frequency domain modes, checking the convergence rates. We further analyse the loss of precision that occurs in this reconstruction due to large pointwise cancellations. We then move on to checking the convergence rates of the $l$-modes of the self-force after subtraction of the regularization parameters. This provides a key check both of our numerical implementation and of the analytical calculation of the regularization parameters for the self-force on eccentric orbits in Kerr spacetime.

Since this is the first calculation of the gravitational self-force on eccentric orbits in Kerr spacetime, there is little possibility of checking our results against the literature. Moreover, since the gravitational self-force is not gauge invariant, we also cannot compare our results --- which are obtained in a certain (completed) radiation gauge --- to results in the circular and/or Schwarzschild limits, where the self-force has only been calculated in the Lorenz and Regge-Wheeler gauges. One thing we can compare with is the average energy and angular momentum fluxes to infinity and down the horizon, which according to the so-called ``balance law'' should be equal to certain orbital averages of the self-force. In Sec.~\ref{sec:blaw}, we will compare these orbit averages to fluxes both from the existing literature and calculated using our own code.

We finally give some sample results from our code. These will be represented as so-called ``self-force loops''  first introduced in \cite{Vega:2013wxa}.

\subsection{Time domain reconstruction}\label{sec:TDrecon}

\begin{figure}
\includegraphics[width=\columnwidth]{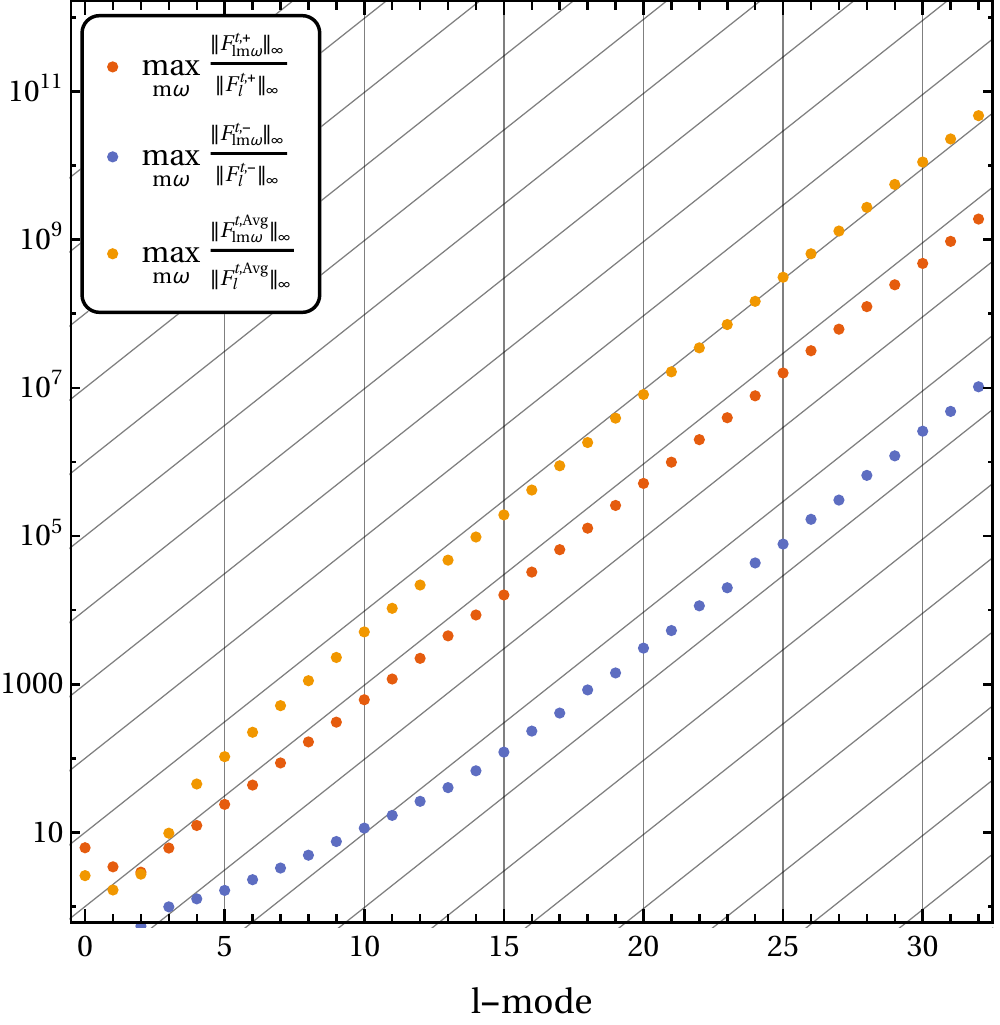}
\caption{Illustration of the loss of precession in construction the $l$-modes from the frequency domain modes as a function of $l$ for an orbit with parameters $(a,p,e)=(0.9,5.5,0.3)$. The loss of precision is measured by taking the maximal value of the sup-norms of all the frequency domain modes $F^t_{lm\omega}$ contributing to a certain $l$-mode $F^t_{l}$ normalized by the sup-norm of the complete $l$-mode. The diagonal gridlines show a reference growth proportional to $(\frac{x_{max}}{x_{min}})^l$ with $x=(r-r_+)$.
}\label{fig:preclossvsl}
\end{figure}
\begin{figure}
\includegraphics[width=\columnwidth]{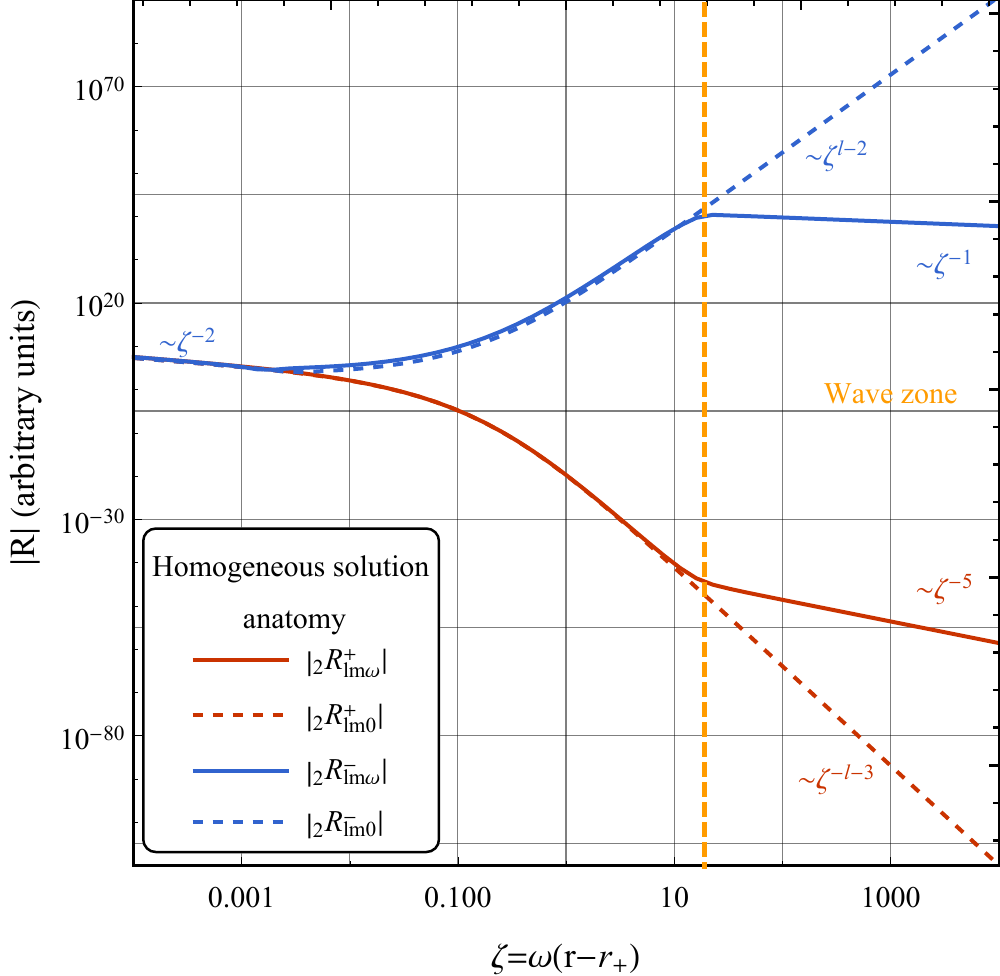}
\caption{The anatomy of a typical homogeneous solution to the $s=+2$ Teukolsky equation. The particular solutions plotted are for $a=0.9$, $l=20$, $m=1$, and $\omega=0.5$ (as solid lines) and $\omega=0$ (as dashed lines). Qualitatively, other solutions look the same. At large radii, the $\omega\neq 0$ modes scale as $z^{-5}$ for the outside modes and $\zeta^{-1}$ for the inside mode. In contrast, the static modes scale as $\zeta^{-l-3}$ and $\zeta^{l-2}$ respectively. Below a radius $\zeta\sim 12$ the oscillating modes become similar to the static modes initially showing the same $\zeta^{-l-3}$ and $\zeta^{l-2}$ behaviour. Near to the horizon all modes scale as $\zeta^{-2}$. (For the sake of this figure all modes have been normalized to numerically agree in this limit.)  
}\label{fig:anatomy}
\end{figure}
\begin{figure}
\includegraphics[width=\columnwidth]{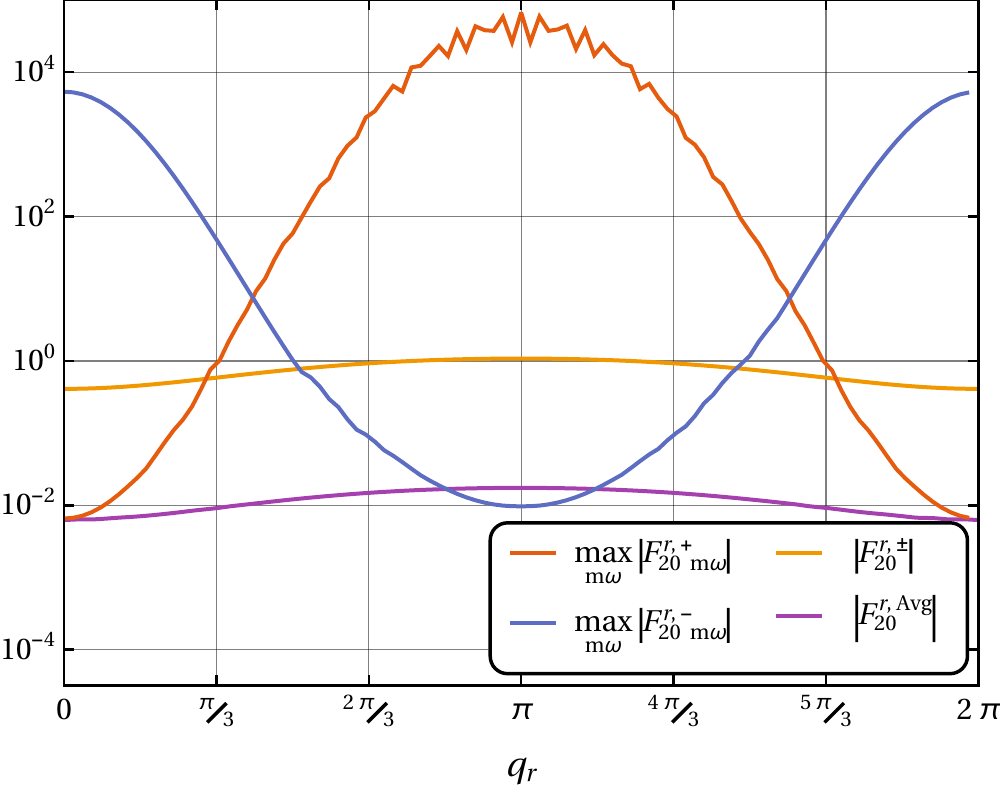}
\caption{
An illustration of the precision loss that occurs when summing over all frequency modes as a function of the orbital phase $q_r$. Shown are both the maximal contributions from the frequency modes $F_{20 m\omega}^{r,\pm}$ to the $l$-mode  $F_{20}^{r,\pm}$, and the final $l$-mode (both the one side values and the two-sided average). The ``$+$''  modes lose most precision at periastron and the ``$-$''  modes lose most precision at apastron. The values are obtained for our standard reference orbit with parameters $(a,p,e)=(0.9,5.5,0.3)$.
}\label{fig:orbitplots}
\end{figure}
As a first test of our implementation we consider the reconstruction of the ``time-domain'' spherical harmonic modes $F_{lm}^{\mu,\Avg}$ for the frequency domain modes $F_{lm\omega}^{\mu,\Avg}$,
\begin{equation}\label{eq:TDrecon}
F_{lm}^{\mu,\Avg}	=\sum_{n=-\infty}^\infty F_{lm\omega_{mn}}^{\mu,\Avg},
\end{equation}
with $\omega_{mn}= m\Omega_\phi +n \Omega_r$.

In our method, we construct $F_{lm\omega}^{\mu,\Avg}$ from the vacuum solutions of $h_{\mu\nu}$ outside of the libration region $\rmin<r<\rmax$ analytically extended to the particle location, rather than the non-vacuum inhomogeneous solutions that would be obtained through variation of parameters. The ``method of extended homogeneous solutions'' was originally introduced in \cite{Barack:2008ms} to avoid poor convergence of the sum over frequency modes due to Gibbs waves caused by the non-differentiability of the retarded field at the particle location. In our method it is doubly necessary because the CCK metric reconstruction procedure is only well-defined for vacuum perturbations.

The expectation of \cite{Barack:2008ms} is that for radial harmonic number $\abs{n}$ large enough, summand of Eq. \eqref{eq:TDrecon} decays exponentially with $\abs{n}$. To test this expectation Fig.~\ref{fig:nconvplot} plots the sup-norm of the frequency modes of $F^{t,\pm}$ normalized by  $\norm{F_{lm}^{\mu,\pm}}_\infty$ for a variety of values for $l$ and $m$ obtained for an eccentric equatorial orbit with parameters $(a,p,e)=(0.9,5.5,0.3)$. (Results for different orbits, values of $l$ and $m$, and components of the self-force are qualitatively similar.) We see that the large-$\abs{n}$ behaviour of the frequency domain modes is consistent with a decay faster than $e^{\abs{n}}$ (shown for reference as the diagonal grid lines), as expected.

Figure~\ref{fig:nconvplot}, however, also highlights a somewhat distressing feature of the time domain reconstruction. We see that a significant fraction of the frequency modes have normalized values which are orders of magnitude larger than~1. This means that there must be significant cancellation between the different frequency modes as they are summed to recover the time domain modes. Consequently, we expect a significant loss of precision as a result. Moreover, as illustrated in Fig.~\ref{fig:preclossvsl} this loss of precision appears to increase exponentially with $l$.

This behaviour is inherent to the method of extended homogeneous solutions. To understand its origin we must take a closer look at the behaviour of the homogeneous solutions of the Teukolsky equation. Figure~\ref{fig:anatomy} shows the anatomy of a set of typical homogeneous solutions to the (spin-2) Teukolsky equation. For values of $r$ large compared to $\omega^{-1}$, the asymptotic wave behaviour dominates and
\begin{align}
\abs{\R[+]{s}{lm\omega}} &\propto r^{-2s-1} +\bigO(r^{-2s}),\\
\abs{\R[-]{s}{lm\omega}} &\propto r^{-1} +\bigO(r^{0}).
\end{align}
However, when $r \lesssim \omega^{-1}$ the solution becomes approximately stationary and is well approximated by the analytically known stationary solutions $\R[\pm]{s}{lm0}$ (see e.g. \cite{vandeMeent:2015lxa}). Near the horizon, for $\zeta=\omega(r-r_+)\ll 1$ these solutions are proportional to $\zeta^{-2}$, whereas for larger values of $\zeta$ they behave as,
\begin{align}
\abs{\R[+]{s}{lm\omega}} &\propto \zeta^{-l-s-1} +\bigO(z^{-l-s}),\\
\abs{\R[-]{s}{lm\omega}} &\propto \zeta^{l-s} +\bigO(\zeta^{l-s+1}).
\end{align}
Consequently, for strong field eccentric orbits the extended homogeneous frequency modes of the self-force with relatively small frequencies will vary significantly in magnitude along the orbit. The magnitude of this variation increases exponentially with $l$, approximately as
\begin{equation}
\hh{\frac{\rmax-r_{+}}{\rmin-r_{+}}}^l.
\end{equation}
On the other hand, for large frequencies $\omega$ the orbit is completely in the ``wave-zone'', and the homogeneous modes will exhibit a variation whose magnitude is independent of $l$. 

Meanwhile the variation of the time-domain modes $F^\mu_{lm}$ along the orbit is controlled by the source, with the magnitude of the variation proportional to $(\rmax/\rmin)^3$, and certainly not growing faster than linearly in $l$. Consequently, it is expected that the magnitude of the low frequency modes is much larger than of the sum over all frequency modes, i.e. the time domain mode $F^\mu_{lm}$.

We finally observe that the loss of precision in reconstruction of the time domain modes is not constant along the orbit. Figure~\ref{fig:orbitplots} combines plots of the maximal value of $\abs{F^{t,\pm}_{lm\omega}}$ obtained for any $m$ and $\omega$ at fixed $l$ as a function of the orbital phase $q_r$ with plots of the total  $l$-mode, both the one-sided values and the average. The difference in magnitude gives an indication of the accuracy lost in summing over all frequency modes. We see that when summing the outside ``$+$'' modes, we lose most accuracy at $q_r=\pi$ (periapsis) while losing no accuracy at $q_r=0$ (apapsis). Conversely, the inside ``$-$'' modes lose most accuracy at apapsis, while losing little precision at periapsis.

This behaviour is again easily understood from the nature of the method of extended homogeneous solutions. Near apapsis the extended homogeneous modes of the field are close to the ``actual'' field that would be obtained through variation of parameters, hence we expect the magnitude of the frequency modes to be similar to the time domain modes without large cancellations. As the modes are analytically extended further into the libration region, the low frequency modes exhibit their anomalous $\sim r^l$ growth leading to very large cancellations at periapsis. The reverse happens with the extended inside modes, they are close to their actual values at periapsis, and grow towards apapsis.

In methods that can obtain the self-force from either the inside or outside field values such as frequency domain Lorenz gauge calculations in Schwarzschild \cite{Akcay:2013wfa,Osburn:2014hoa} or scalar field calculations in Kerr \cite{Warburton:2011hp}, this orbital phase dependence of the accuracy loss offers an easy way to mitigate its impact. One simply uses the field on the side of the particle that exhibits the least precision loss. Unfortunately, the ``no-string'' radiation gauge procedure used in this paper needs the average of the field on both sides of the orbit. Hence we have no other option than to knuckle up and bear the loss of accuracy. Luckily, since our code is implemented using arbitrary precision arithmetic it is straightforward to simply ask for more precision (at the cost of computation time).

Currently, this precision loss in the summation of the frequency modes, appears to be the main limiting factor in pushing our calculation to higher eccentricities and accuracies. For this reason, we have dwelled on this phenomenon to some length above. Our main conclusions are as follows, the impact of this effect is greatest for:
\begin{itemize}
\item Orbits with high eccentricities.
\item Orbits in the strong field domain (which have more modes for which the stationary ``near-zone'' domain envelops the orbit).
\item High $l$ modes.
\item Low frequency modes.
\end{itemize}
The third point is exacerbated in Kerr spacetime by the fact that the spherical $l$-modes needed to calculate the self-force in our mode-sum scheme are combinations of spheroidal $\spl$-modes, including modes with $\spl>l$. However, this is at least mitigated by the last conclusion, since low $\omega$ modes exhibit only a small spread in $\spl$-modes contributing to a given $l$-mode.

A solid understanding of this phenomenon and its causes will also allow us to better control its impact in future evolutions of our code. For example, the fact that only low frequency modes are affected is good news. The nature of the MST methods being used means that is much easier to generate more precision for low frequency modes than it is for high frequency modes. The same is true for other numerical steps such as the integrals needed to calculate the mode amplitudes $Z_{lm\omega}^\pm$. Consequently, by fine tuning the precision requested for each mode we can minimize the impact of this phenomenon on computation time at fixed requested overall precision as we increase eccentricity. Currently, we have taken the somewhat ham-handed approach of simply increasing the overall requested precision to mitigate the precision loss, leading to a lot of wasted resources on obtaining precision for modes that will not add to the precision of the overall result.

\subsection{Regularization parameters}
\begin{figure}
\includegraphics[width=\columnwidth]{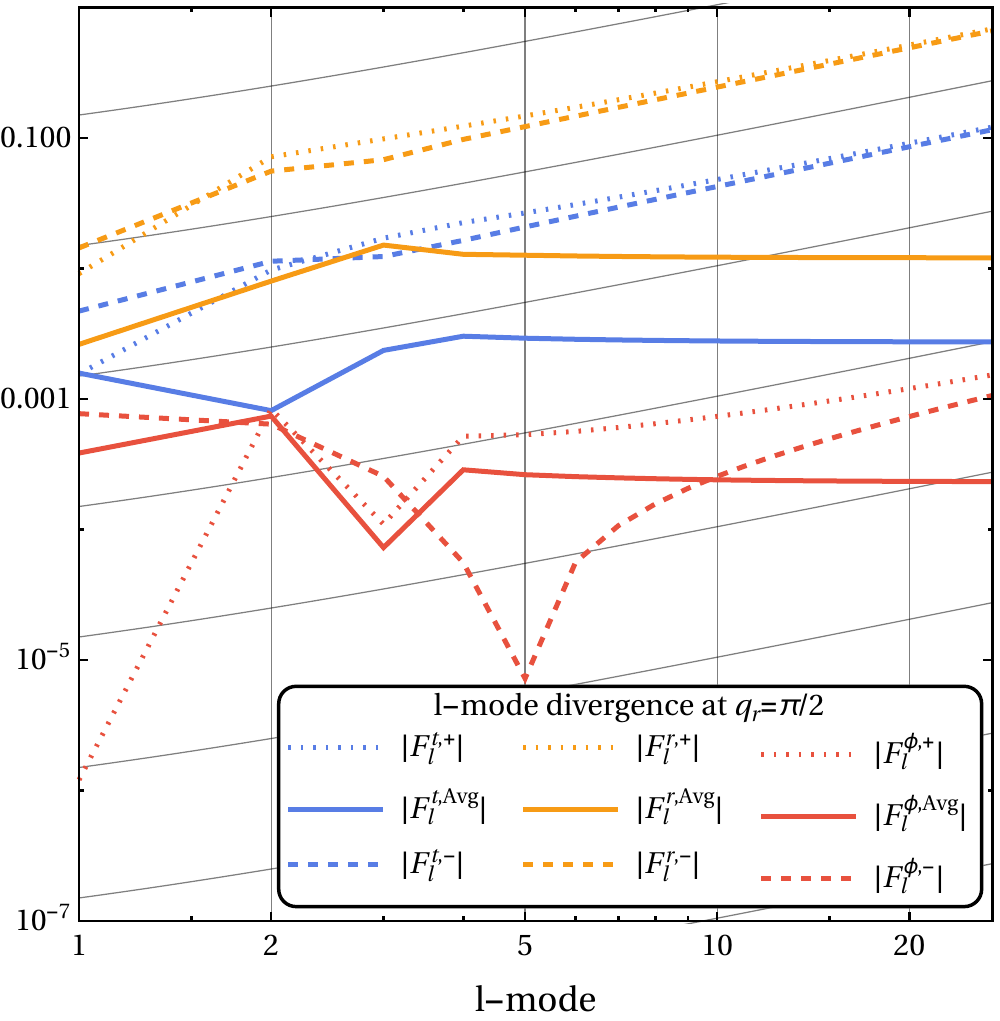}
\caption{The $l$-modes of the self-force on an orbit with parameters $(a,p,e)=(0.9,5.5,0.3)$ (before regularization) at a generic point $q_r=\pi/2$ along the orbit. The inside and outside values follow the expected $\bigO(l)$ growth (indicated by the diagonal gridlines), while the two-sided averages asymptote to a constant value. }\label{fig:regplot0}
\end{figure}
\begin{figure}
\includegraphics[width=\columnwidth]{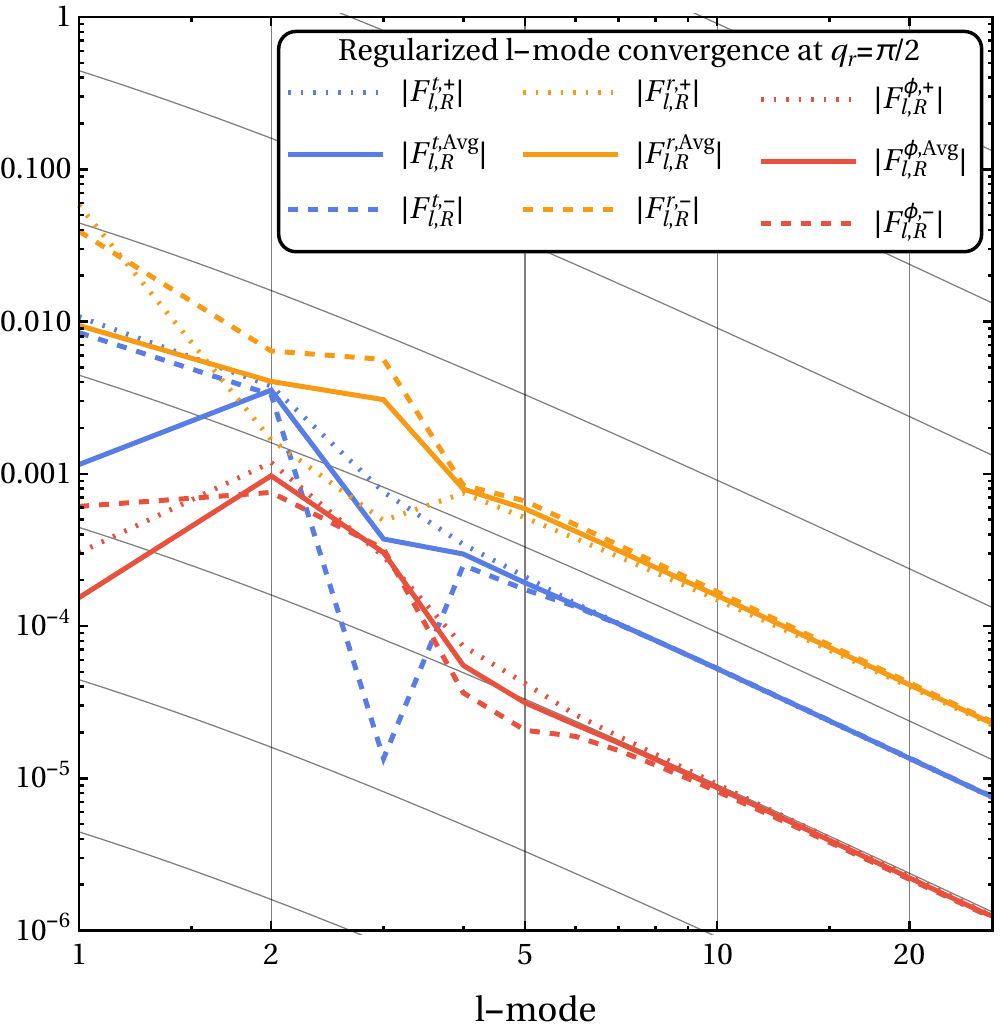}
\caption{The same $l$-modes as in Fig.~\ref{fig:regplot0} after subtracting the regularization parameters. All components follow the expected $\bigO(l^{-2})$ behaviour.}\label{fig:regplot2}
\end{figure}
\begin{figure}
\includegraphics[width=\columnwidth]{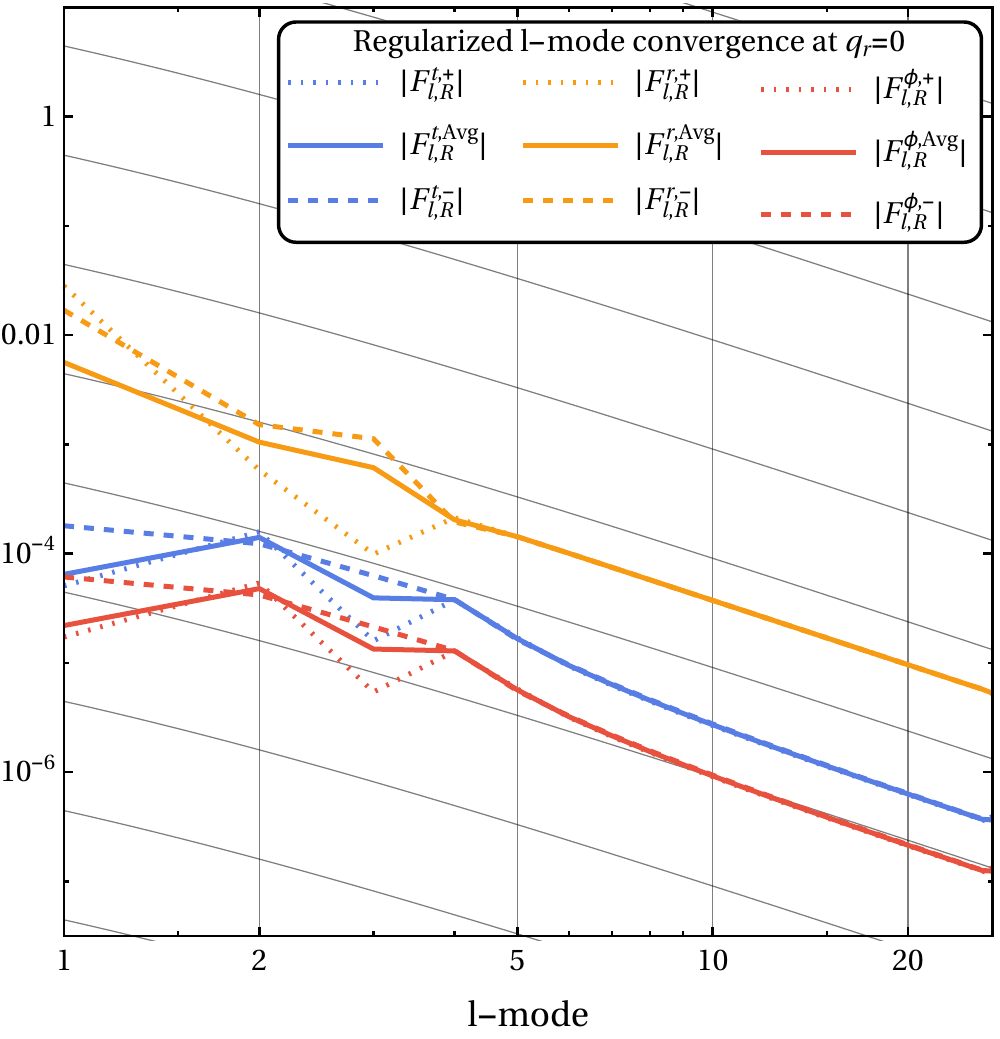}
\caption{
Convergence of the (regularized) $l$-modes at  apapsis ($q_r=0$). Despite the $t$ and $\phi$ components of the self-force being regular at this point, the $l$-mode decay only as $\bigO(l^{-2})$ due to the non-smoothness of the extension.
}\label{fig:regplot1}
\end{figure}
\begin{figure}
\includegraphics[width=\columnwidth]{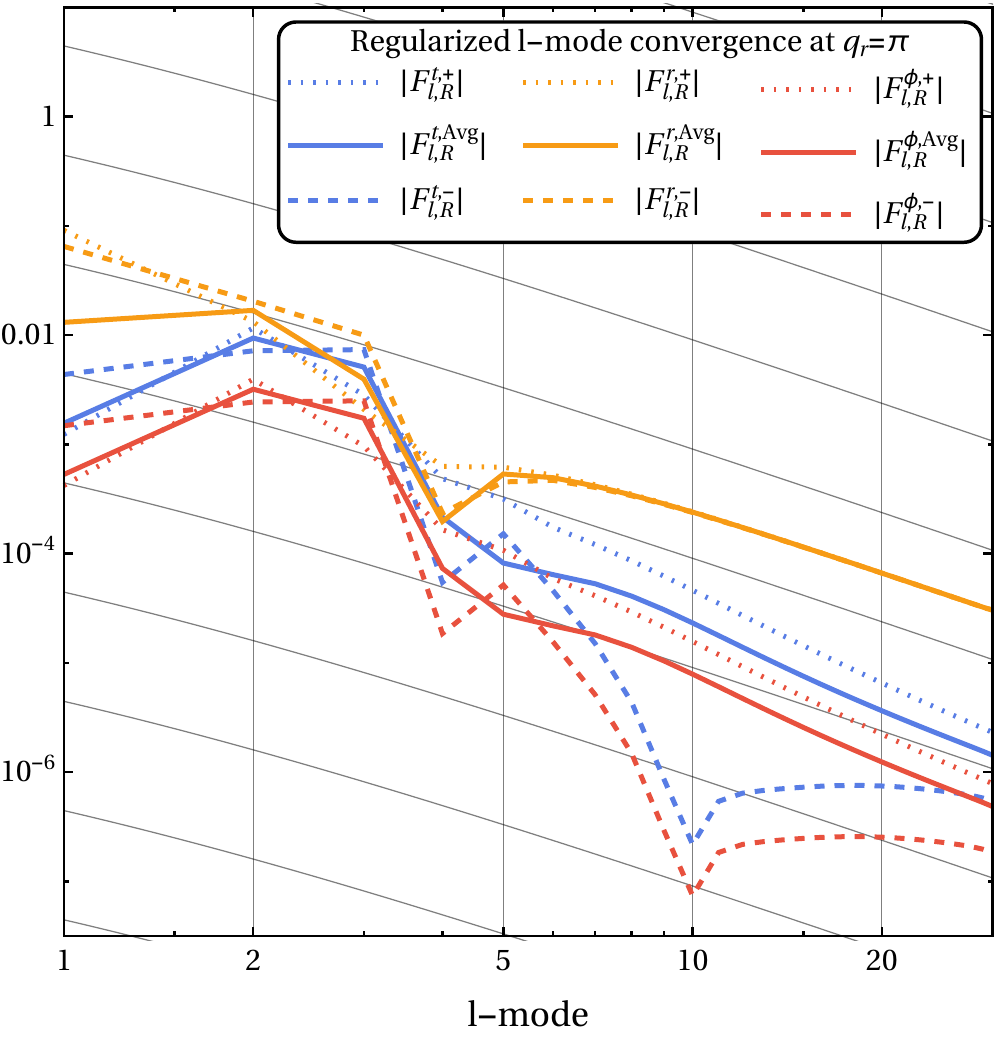}
\caption{
Similar, to Fig.~\ref{fig:regplot1} but now evaluated at periapsis ($q_r=\pi$). Again we see that the $t$ and $\phi$ components show convergence no faster than $\bigO(l^{-2})$ due to the non-smoothness of the extension.
}\label{fig:regplot3}
\end{figure}


For our next test we will compare the $l$-modes $F_l^{\mu}$ obtained numerical through our procedure to the regularization parameters $A^\mu$, $B^\mu$, and $C^\mu$ first derived in \cite{Barack:2002mh} from the singular field in Lorenz gauge. According to the analysis of \cite{Pound:2013faa}, the $l$-modes obtained through the radiation gauge procedure utilized here should satisfy
\begin{equation}
F_l^{\mu,\pm} = \pm (l+1/2) A^\mu + B^\mu +\frac{C^\mu}{l+1/2} +\bigO(l^{-2}), 
\end{equation}
for the one-sided values, and
\begin{equation}
F_l^{\mu,\Avg} =  B^\mu +\frac{C^\mu}{l+1/2} +\bigO(l^{-2}), 
\end{equation}
for the two-sided average with the regularization parameters taking their Lorenz gauge values.

Comparing our numerical results to the analytic expectations provides a crucial check of our numerical procedure for calucalting the self-force in Kerr spacetime. Almost any implementation error (crucially including overall sign errors) would cause our calculated field to fail to match the predicted large $l$ behaviour. Additionally, this test will provide a first numerical validation of the highly non-trivial analytical calculation of \cite{Barack:2002mh} and analysis of \cite{Pound:2013faa} for eccentric equatorial orbits in Kerr spacetime.

In Fig.~\ref{fig:regplot0}, we plot the bare values of the $l$-modes evaluated at a generic point (say $q_r=\pi/2$) along the orbit obtained numerically for a reference orbit with parameters $(a,p,e)=(0.9,5.5,0.3)$. As expected, the one-sided values of the individual components of the self-force show a $\bigO(l)$  divergence at large-$l$, while the two-sided averages asymptote to a constant value. Note how the one-sided values of the $\phi$ component have not yet reached their asymptotic behaviour at the largest $l$ included in the plot, whereas the two-sided average of the same component is already much better behaved.

Next we calculate the ``regularized'' $l$-modes $F^\mu_{l,\reg}$ defined by 
\begin{equation}\label{eq:regF2}
F_{l,\reg}^{\mu,\pm} = F_{l}^{\mu,\pm} \mp (l+1/2)A^\mu - B^\mu - \frac{C^\mu}{l+1/2},
\end{equation}
for the one-sided values and by Eq.~\eqref{eq:regF} for the two-sided average.

Figure~\ref{fig:regplot2} shows the regularized $l$-modes for the same orbital parameters $(a,p,e)=(0.9,5.5,0.3)$ and phase $q_r=\pi/2$ as used in Fig.~\ref{fig:regplot0}. We see that for all three non-zero components of the self-force the inside, outside and average values of the regularized $l$-modes neatly exhibit the expected $\bigO(l^{-2})$ fall-off at large-$l$. This means that the Lorenz gauge regularization parameters are spot on for our calculation, providing a huge boost in confidence in both the method and numerical implementation.

In Fig.~\ref{fig:regplot1} and \ref{fig:regplot3} we finally look at the (regularized) $l$-modes at two special points along the orbit, the periapsis and apapsis. At these points the $t$ and $\phi$-components of the self-force are expected to be regular, in particular  the regularization parameters for these components all vanish at these points. Naively, one would therefore expect the $l$-modes to fall-off exponentially at large $l$. However, looking at Figs.~\ref{fig:regplot1} and \ref{fig:regplot3} we see that this is not the case. Instead the $t$ and $\phi$ components show the same $\bigO(l^{-2})$ behaviour at large $l$ as the other regularized modes. This can be understood as a consequence of the extension of the full self-force $\Fext^\mu$ constructed in Sec.~\ref{sec:GSF}, which is just smooth enough to keep the regularization parameters unchanged. This observation underlines the fact that the ``higher-order regularization'' parameters of our radiation gauge calculation should not (and do not) match the ones known from Lorenz calculations \cite{Heffernan:2012su,Heffernan:2012vj}.

\subsection{Balance law}\label{sec:blaw}
\begingroup
\squeezetable
\begin{table*}[P]
\caption{Numerical test of the balance law for a selection of strong field orbits. In each entry the first row gives $\mr^{-1}(\langle\d{\mathcal{C}_\mathrm{flux}^\pI}{t}\rangle + \langle\d{\mathcal{C}_\mathrm{flux}^\pH}{t}\rangle)$ (with $\mathcal{C}$ either $\nE$ or $\nL$) calculated from the asymptotic values of $\psi_4$. The second row gives $-\mr^{-1}\langle\d{\mathcal{C}^\mathrm{GSF}}{t}\rangle$. These independently calculated quantities agree upto the estimated error level, providing a strong consistency check of the radiation self-force formalism, our numerical implementation, and error estimates. The brackets $(.)$ at the end of values indicated the estimated uncertainty on the last digit(s) (e.g. $ 1.240349(2)\times 10^{-4}$ indicates $ 1.240349\times 10^{-4} \pm 2\times 10^{-10}$).
}\label{tab:blaw}
\[\begin{array}{d{3}d{3}d{3}|lr|lr}
\toprule
 a & p & e & \mr^{-1}\avg{\d{\nE}{t}} & \text{rel. diff.} & \mr^{-1}\avg{\d{\nL}{t}} & \text{rel. diff.}  \Ts\Bs\\
\hline
 -0.99 & 9.5 & 0.1 & 
\begin{array}[t]{l}
 1.240352212605(5)\times 10^{-4} \\
 1.240349(2)\times 10^{-4} \\
\end{array}
 & -2.9\times 10^{-6} & 
\begin{array}[t]{l}
 3.35399692067(1)\times 10^{-3} \\
 3.354001(7)\times 10^{-3} \\
\end{array}
 & 1.3\times 10^{-6}\Ts\\
\hline
 -0.99 & 11. & 0.1 & 
\begin{array}[t]{l}
 5.02889013411(6)\times 10^{-5} \\
 5.02890(1)\times 10^{-5} \\
\end{array}
 & 2.3\times 10^{-6} & 
\begin{array}[t]{l}
 1.73631631341(2)\times 10^{-3} \\
 1.736319(2)\times 10^{-3} \\
\end{array}
 & 1.5\times 10^{-6}\Ts\\
\hline
 -0.99 & 9.7 & 0.2 & 
\begin{array}[t]{l}
 1.426974820(5)\times 10^{-4} \\
 1.426985(7)\times 10^{-4} \\
\end{array}
 & 7.1\times 10^{-6} & 
\begin{array}[t]{l}
 3.59197541(1)\times 10^{-3} \\
 3.59201(1)\times 10^{-3} \\
\end{array}
 & 9.2\times 10^{-6}\Ts\\
\hline
 -0.99 & 11. & 0.2 & 
\begin{array}[t]{l}
 5.68947089758(5)\times 10^{-5} \\
 5.68947(3)\times 10^{-5} \\
\end{array}
 & 2.7\times 10^{-7} & 
\begin{array}[t]{l}
 1.82140420957(1)\times 10^{-3} \\
 1.821403(5)\times 10^{-3} \\
\end{array}
 & -5.0\times 10^{-7}\Ts\\
\hline
 -0.99 & 10. & 0.3 & 
\begin{array}[t]{l}
 1.526199(2)\times 10^{-4} \\
 1.526216(9)\times 10^{-4} \\
\end{array}
 & 1.1\times 10^{-5} & 
\begin{array}[t]{l}
 3.602085(3)\times 10^{-3} \\
 3.60213(1)\times 10^{-3} \\
\end{array}
 & 1.2\times 10^{-5}\Ts\\
\hline
 -0.99 & 11. & 0.3 & 
\begin{array}[t]{l}
 6.82322768(4)\times 10^{-5} \\
 6.82320(4)\times 10^{-5} \\
\end{array}
 & -4.0\times 10^{-6} & 
\begin{array}[t]{l}
 1.962751534(10)\times 10^{-3} \\
 1.96274(1)\times 10^{-3} \\
\end{array}
 & -4.8\times 10^{-6}\Ts\\
\hline
 -0.99 & 10.3 & 0.4 & 
\begin{array}[t]{l}
 1.60866(2)\times 10^{-4} \\
 1.60874(8)\times 10^{-4} \\
\end{array}
 & 4.9\times 10^{-5} & 
\begin{array}[t]{l}
 3.54613(4)\times 10^{-3} \\
 3.5463(2)\times 10^{-3} \\
\end{array}
 & 5.2\times 10^{-5}\Ts\\
\hline
 -0.99 & 11. & 0.4 & 
\begin{array}[t]{l}
 8.479613(4)\times 10^{-5} \\
 8.4797(1)\times 10^{-5} \\
\end{array}
 & 8.7\times 10^{-6} & 
\begin{array}[t]{l}
 2.1595054(8)\times 10^{-3} \\
 2.15951(3)\times 10^{-3} \\
\end{array}
 & 3.8\times 10^{-6} \Ts\\
\hline
 0.5 & 5. & 0.1 & 
\begin{array}[t]{l}
 1.8133382543991(9)\times 10^{-3} \\
 1.81333(2)\times 10^{-3} \\
\end{array}
 & -4.8\times 10^{-6} & 
\begin{array}[t]{l}
 2.062659697674(1)\times 10^{-2} \\
 2.06265(3)\times 10^{-2} \\
\end{array}
 & -6.7\times 10^{-6} \Ts\\
\hline
 0.5 & 6. & 0.1 & 
\begin{array}[t]{l}
 7.093793531283(8)\times 10^{-4} \\
 7.09374(6)\times 10^{-4} \\
\end{array}
 & -6.9\times 10^{-6} & 
\begin{array}[t]{l}
 1.053488681053(1)\times 10^{-2} \\
 1.05348(1)\times 10^{-2} \\
\end{array}
 & -5.4\times 10^{-6} \Ts\\
\hline
 0.5 & 5. & 0.2 & 
\begin{array}[t]{l}
 2.0871627012(8)\times 10^{-3} \\
 2.08713(10)\times 10^{-3} \\
\end{array}
 & -1.5\times 10^{-5} & 
\begin{array}[t]{l}
 2.2076791923(7)\times 10^{-2} \\
 2.2076(1)\times 10^{-2} \\
\end{array}
 & -3.3\times 10^{-5} \Ts\\
\hline
 0.5 & 6. & 0.2 & 
\begin{array}[t]{l}
 7.77122991658(2)\times 10^{-4} \\
 7.7711(2)\times 10^{-4} \\
\end{array}
 & -1.8\times 10^{-5} & 
\begin{array}[t]{l}
 1.082908213382(3)\times 10^{-2} \\
 1.08290(3)\times 10^{-2} \\
\end{array}
 & -7.8\times 10^{-6} \Ts\\
\hline
 0.5 & 5. & 0.3 & 
\begin{array}[t]{l}
 2.6006571(2)\times 10^{-3} \\
 2.6005(2)\times 10^{-3} \\
\end{array}
 & -4.4\times 10^{-5} & 
\begin{array}[t]{l}
 2.4798414(2)\times 10^{-2} \\
 2.4797(2)\times 10^{-2} \\
\end{array}
 & -6.0\times 10^{-5} \Ts\\
\hline
 0.5 & 6. & 0.3 & 
\begin{array}[t]{l}
 8.86676911(8)\times 10^{-4} \\
 8.8666(5)\times 10^{-4} \\
\end{array}
 & -2.3\times 10^{-5} & 
\begin{array}[t]{l}
 1.127740300(8)\times 10^{-2} \\
 1.12772(5)\times 10^{-2} \\
\end{array}
 & -2.1\times 10^{-5} \Ts\\
\hline
 0.5 & 5. & 0.4 & 
\begin{array}[t]{l}
 3.53058(2)\times 10^{-3} \\
 3.528(1)\times 10^{-3} \\
\end{array}
 & -7.4\times 10^{-4} & 
\begin{array}[t]{l}
 2.97986(2)\times 10^{-2} \\
 2.9779(8)\times 10^{-2} \\
\end{array}
 & -6.7\times 10^{-4} \Ts\\
\hline
 0.5 & 6. & 0.4 & 
\begin{array}[t]{l}
 1.0309895(6)\times 10^{-3} \\
 1.03097(7)\times 10^{-3} \\
\end{array}
 & -1.8\times 10^{-5} & 
\begin{array}[t]{l}
 1.1805233(6)\times 10^{-2} \\
 1.1805(1)\times 10^{-2} \\
\end{array}
 & -1.9\times 10^{-5}\Ts\\
\hline
 0.99 & 2. & 0.1 & 
\begin{array}[t]{l}
 4.4073701(1)\times 10^{-2} \\
 4.40(1)\times 10^{-2} \\
\end{array}
 & -2.5\times 10^{-3} & 
\begin{array}[t]{l}
 1.65690967(5)\times 10^{-1} \\
 1.653(6)\times 10^{-1} \\
\end{array}
 & -2.3\times 10^{-3}\Ts\\
\hline
 0.99 & 3. & 0.1 & 
\begin{array}[t]{l}
 1.08256949688(3)\times 10^{-2} \\
 1.0819(2)\times 10^{-2} \\
\end{array}
 & -6.0\times 10^{-4} & 
\begin{array}[t]{l}
 6.5830999430(2)\times 10^{-2} \\
 6.579(2)\times 10^{-2} \\
\end{array}
 & -6.2\times 10^{-4}\Ts\\
\hline
 0.99 & 2. & 0.2 & 
\begin{array}[t]{l}
 4.7242644(7)\times 10^{-2} \\
 4.69(2)\times 10^{-2} \\
\end{array}
 & -7.2\times 10^{-3} & 
\begin{array}[t]{l}
 1.7000999(2)\times 10^{-1} \\
 1.688(8)\times 10^{-1} \\
\end{array}
 & -7.0\times 10^{-3}\Ts\\
\hline
 0.99 & 3. & 0.2 & 
\begin{array}[t]{l}
 1.1530343191(3)\times 10^{-2} \\
 1.1535(7)\times 10^{-2} \\
\end{array}
 & 3.7\times 10^{-4} & 
\begin{array}[t]{l}
 6.683744156(1)\times 10^{-2} \\
 6.685(7)\times 10^{-2} \\
\end{array}
 & 1.9\times 10^{-4}\Ts\\
\hline
 0.99 & 2. & 0.3 & 
\begin{array}[t]{l}
 5.250991(4)\times 10^{-2} \\
 5.22(6)\times 10^{-2} \\
\end{array}
 & -6.6\times 10^{-3} & 
\begin{array}[t]{l}
 1.771962(1)\times 10^{-1} \\
 1.76(2)\times 10^{-1} \\
\end{array}
 & -5.8\times 10^{-3}\Ts\\
\hline
 0.99 & 3. & 0.3 & 
\begin{array}[t]{l}
 1.26252561(5)\times 10^{-2} \\
 1.265(2)\times 10^{-2} \\
\end{array}
 & 1.8\times 10^{-3} & 
\begin{array}[t]{l}
 6.8247192(3)\times 10^{-2} \\
 6.84(2)\times 10^{-2} \\
\end{array}
 & 1.8\times 10^{-3}\Ts\\
\hline
 0.99 & 2. & 0.4 & 
\begin{array}[t]{l}
 5.99553(2)\times 10^{-2} \\
 6.10(9)\times 10^{-2} \\
\end{array}
 & 1.7\times 10^{-2} & 
\begin{array}[t]{l}
 1.874314(6)\times 10^{-1} \\
 1.90(3)\times 10^{-1} \\
\end{array}
 & 1.6\times 10^{-2}\Ts\\
\hline
 0.99 & 3. & 0.4 & 
\begin{array}[t]{l}
 1.397344(1)\times 10^{-2} \\
 1.397(6)\times 10^{-2} \\
\end{array}
 & -5.6\times 10^{-4} & 
\begin{array}[t]{l}
 6.960990(5)\times 10^{-2} \\
 6.96(4)\times 10^{-2} \\
\end{array}
 & -1.3\times 10^{-4} \\
\botrule
\end{array}\]

\end{table*}
\endgroup
In this section we compare two independent ways of calculating the (long-term) change in the energy and angular momentum of a particle in orbit around a black hole through interaction with its own gravitational field. This change can be obtained either from the force acting on the particle or by monitoring the total energy and angular momentum leaving the system through future null infinity and the future horizon of the black hole. 

Starting from the definitions of the specific energy and angular momentum,
\begin{align}
\nE &= -u^\mu \hh{\pd{}{t}}_\mu,\\
\nL &= u^\mu\hh{\pd{}{\phi}}_\mu,
\end{align}
their average rate of change over an orbital period is obtained by differentiating with respect to $\pt$, substituting Eq. \eqref{eq:gdGSF}, and integrating over an orbital period $\mathcal{T}_r$,
\begin{align}
\avg{\d{\nE^\mathrm{GSF}}{t}} &= \frac{\mr}{T_r} \int_0^{\mathcal{T}_r} -F_t(\pt) \id{\pt}\text{, and}\\
\avg{\d{\nL^\mathrm{GSF}}{t}} &= \frac{\mr}{T_r} \int_0^{\mathcal{T}_r} F_\phi(\pt) \id{\pt}.
\end{align}

On the other hand, we can also obtain the average flux of energy and angular momentum to future null-infinity and down the future black hole horizon directly from the frequency domain solutions of the Teukolsky equation for $\psi_4$. The fluxes at infinity can be extracted straightforwardly,
\begin{align}
\avg{\d{\nE_\mathrm{flux}^\pI}{t}} &=
	\frac{\mr}{4\pi} \sum_{\spl m\omega} \frac{\abs{Z^{\pI}_{\spl m\omega}}^2}{\omega^2},\\
\avg{\d{\nL_\mathrm{flux}^\pI}{t}} &=
	\frac{\mr}{4\pi} \sum_{\spl m\omega} \frac{\abs{Z^{\pI}_{\spl m\omega}}^2}{\omega^3}.
\end{align}
With a little more work Teukolsky and Press \cite{Teukolsky:1974yv} showed how to extract the horizon fluxes as well,
\begin{align}
\avg{\d{\nE_\mathrm{flux}^\pH}{t}} &=
	\frac{\mr}{4\pi} \sum_{\spl m\omega} p_{\spl m\omega} \frac{\abs{Z^{\pH}_{\spl m\omega}}^2}{\omega^2},\\
\avg{\d{\nL_\mathrm{flux}^\pH}{t}} &=
	\frac{\mr}{4\pi} \sum_{\spl m\omega} p_{\spl m\omega} \frac{\abs{Z^{\pH}_{\spl m\omega}}^2}{\omega^3},
\end{align}
where $p_{\spl m\omega}$ is the Teukolsky-Starobinsky constant defined in Eq. \eqref{eq:TSconstant}.

It was shown by Mino \cite{Mino:2003yg,Mino:2005an,Mino:2005yw} (reproducing some earlier (partial) results of Quinn and Wald \cite{Quinn:1999kj} and Gal'tsov \cite{Gal'tsov:1982zz}) that these changes in energy and angular momentum satisfy the so-called \emph{balance law},
\begin{align}
\avg{\d{\nE^\mathrm{GSF}}{t}} +\avg{\d{\nE_\mathrm{flux}^\pI}{t}} + \avg{\d{\nE_\mathrm{flux}^\pH}{t}} &= 0,\\
\avg{\d{\nL^\mathrm{GSF}}{t}}  +\avg{\d{\nL_\mathrm{flux}^\pI}{t}} + \avg{\d{\nL_\mathrm{flux}^\pH}{t}} &= 0.
\end{align}
That is, the average rate of change of (local) orbital energy and angular momentum is equal to the average rate at which energy and angular momentum are dissipated from the system in gravitational waves.

With our code we can calculate both the local change of energy and angular momentum due to the self-force acting on the particle and the energy and angular momentum fluxes leaving the system for eccentric equatorial orbits in Kerr spacetime. Table \ref{tab:blaw} compares the results from both calculations for a selection of strong field eccentric orbits (with whirl numbers $\Omega_\phi/\Omega_r$ ranging between $2$ and $5$). In all cases the observed differences are comparable to the estimated errors for the result. This not only provides us with a strong consistency check on our results, it also tells us that our estimation of the errors in the numerical result is fairly accurate.

We also see sharp reduction in the obtained precision for strong field orbits with high eccentricity, completely inline with the expectations from Sec.~\ref{sec:TDrecon}.

\subsection{Sample results: Self force loops}
\begin{figure}
\includegraphics[width=\columnwidth]{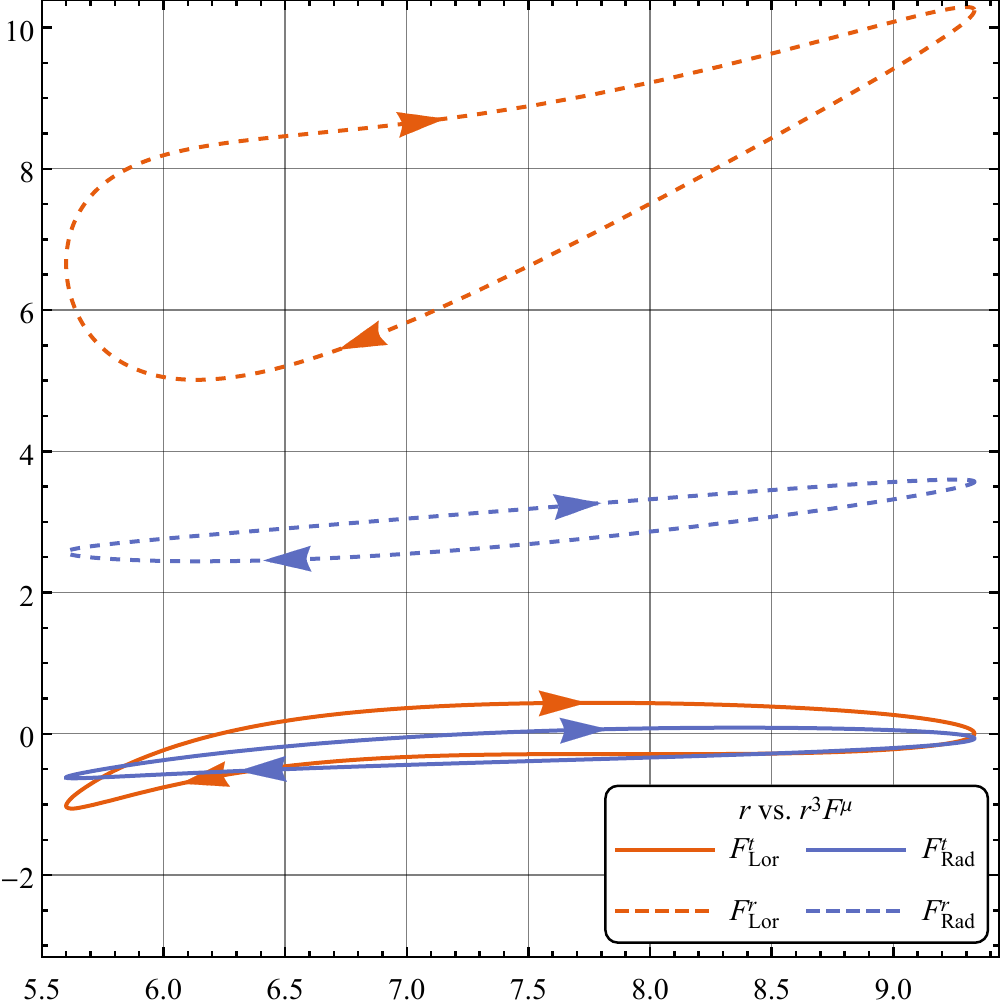}
\caption{
Self-force loops for an orbit with parameters $(a,p,e)=(0,7,0.25)$ calculated in two different gauges using the radiation gauge method of the paper and the Lorenz gauge code of \cite{Akcay:2013wfa}. We find partial overlap in the loops for $F^t$ as is necessitated by the balance law. However, the $F^r$ loops for different gauges are completely disjoint, stressing the gauge dependence of the self-force.
}
\label{fig:compplot}
\end{figure}
\begin{figure*}[P]
\includegraphics[width=\textwidth]{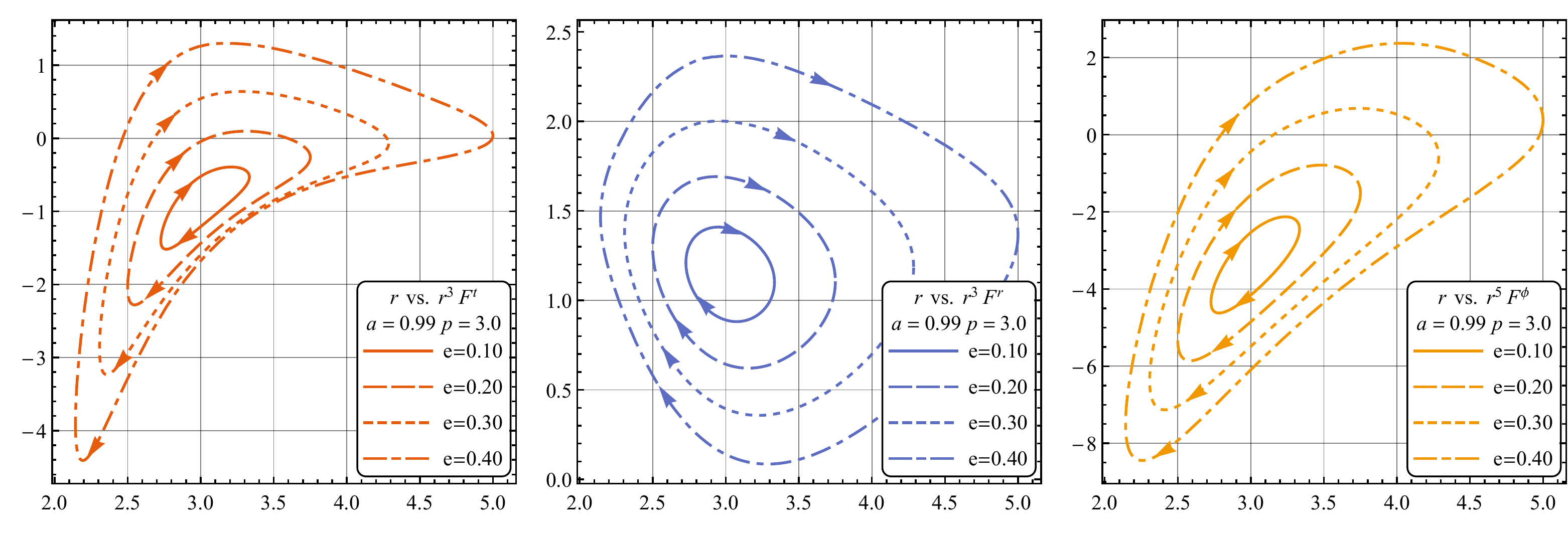}
\includegraphics[width=\textwidth]{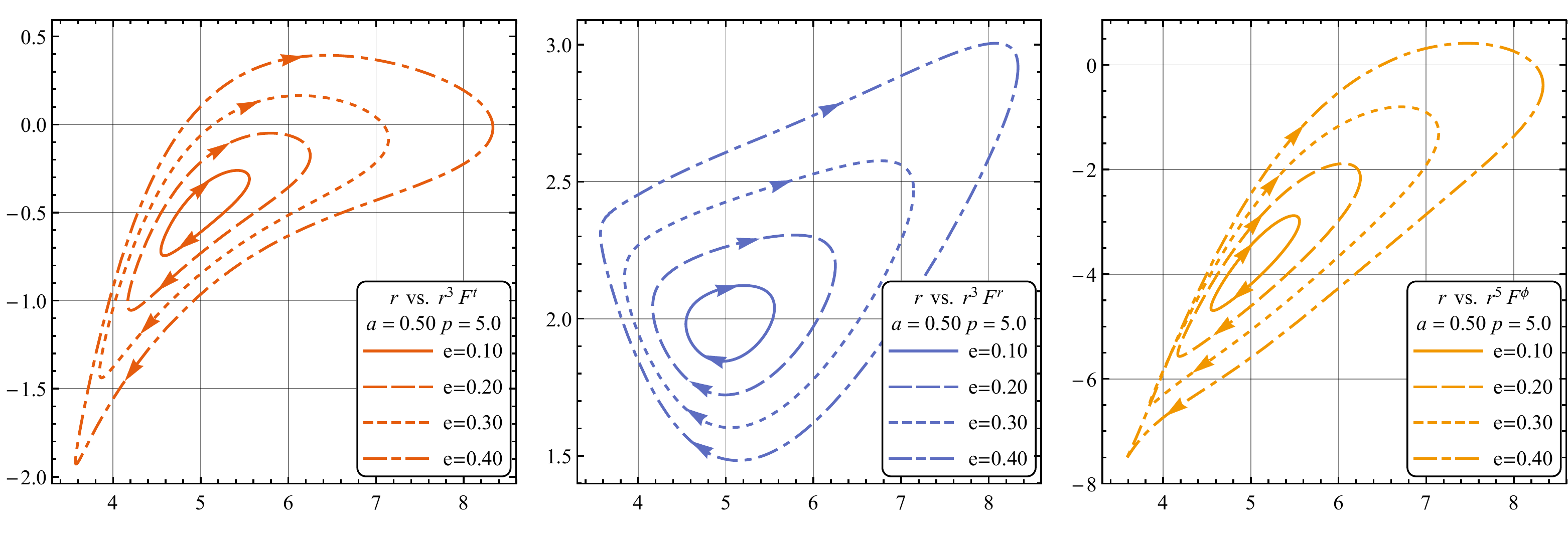}
\includegraphics[width=\textwidth]{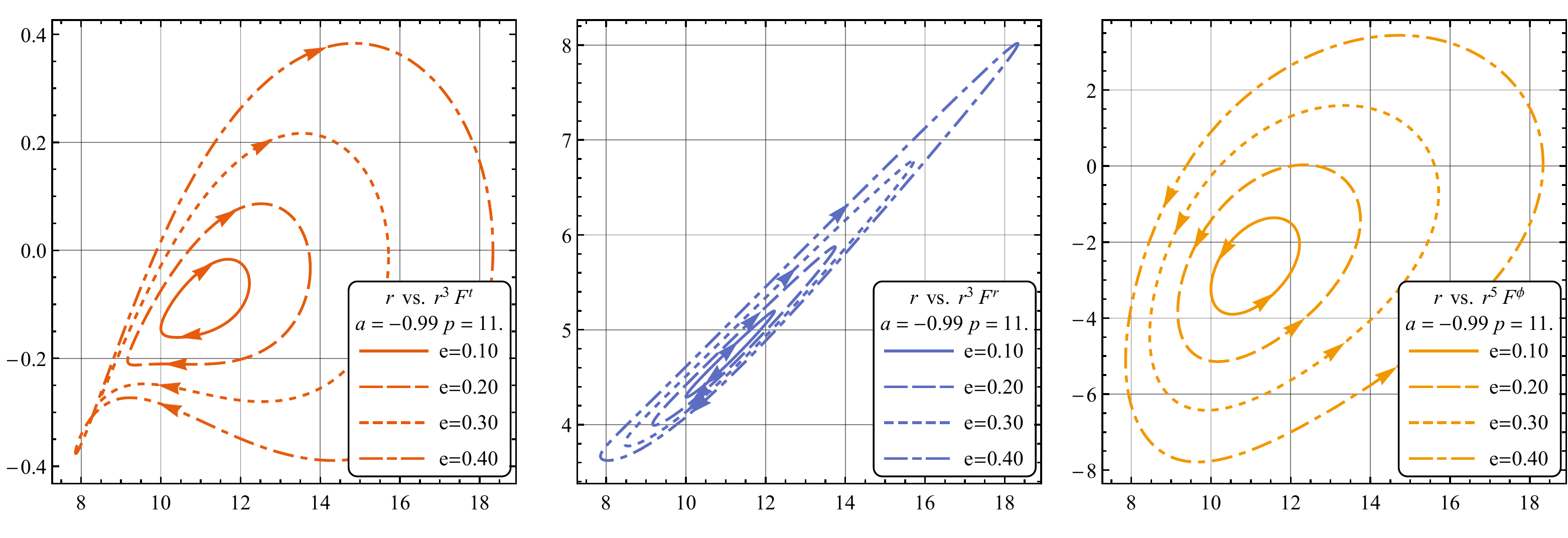}
\caption{``Self-force loops'' for a variety of orbits. On each plot the horizontal axis shows the radial position $r$ of the particle, and the vertical axis shows a component of the self-force, rescaled by an appropriate power of $r$ ($r^3$ for $F^t$ and $F^r$, and $r^5$ for $F^\phi$). Each column shows one component of the self-force ($F^t$ on the left $F^r$ in the middle, and $F^\phi$ on the right), while each row shows orbits with a fixed spin $a$ and semilatus rectum $p$. Complete data for these plots is available as supplementary data \cite{SD2}.}\label{fig:loopplots}
\end{figure*}
We conclude this section by presenting some sample results of the gravitational self-force on a selection of strong field eccentric equatorial orbits. We present the result in so-called ``self-force loops'' introduced by \cite{Vega:2013wxa}. In self-force loop plots, the self-force on an eccentric equatorial orbit is plotted against the radial position $r$ while factoring out the dominant radial scaling (i.e. $r^{-3}$ for $F^t$ and $F^r$ and $r^{-5}$ for $F^\phi$), since the self-force is generally different on the inward part of the orbit than on the outward leg this produces a loop shaped graph.

In Fig. \ref{fig:loopplots} we show self-force loops for a variety of orbital parameters. Each plot shows the self-force loops for a fixed component of the self-force, spin $a$, and semilatus rectum $p$, while varying the eccentricity $e$. For the most part the plots show similar features.

For example, most loops have a clockwise orientation, meaning that the self-force on the outgoing leg is larger than on the ingoing leg. In particular the self-force obtains its maximal value after passing through periapsis. Notable exceptions to this behaviour are found in the $F^\phi$ loops around a retrograde spinning ($a=-0.99$) black hole, which are all anti-clockwise. We further note that the orbit with $(a,p,e)=(-0.99,11,0.4)$ forms a figure eight reversing its orientation near periapsis.

However, we need to remember that there is limited physics in the specific pointwise features of the self-force, since the self-force is not gauge invariant. As a reminder of this fact we have included a plot of the gravitational self-force on an orbit around a Schwarschild black hole with $(p,e)=(7,0.25)$, calculated both using the radiation gauge techniques of this paper and using the Lorenz gauge techniques of \cite{Akcay:2013wfa} in Fig.~\ref{fig:compplot}. The $F^t$ components show overlap (as their average behaviour is dictated by the balance law), but the $F^r$-loops are completely disjoint. This does \emph{not} mean that the gravitational self-force is devoid of physical meaning, instead it means we should consider physical (gauge invariant) observables when comparing results. The orbital averages of $F_t$ and $F_\phi$ discussed in Sec.~\ref{sec:blaw} are examples of gauge invariants constructed from the self-force. Other examples include the redshift invariant calculated in our previous paper \cite{vandeMeent:2015lxa}, the ``self-torque'' exerted on a test spin \cite{Dolan:2013roa}, and tidal invariants \cite{Dolan:2014pja}.

\section{Conclusions and Discussion}\label{sec:discussion}
This paper has provided the first calculation of the gravitational self-force on eccentric equatorial orbits in Kerr spacetime, extending the previous results for the redshift invariant obtained in \cite{vandeMeent:2015lxa}. Our method employs the Chrzanowski-Cohen-Kegeles formalism to reconstruct the local radiation gauge metric perturbation from the Weyl scalar $\psi_4$, and then applys the results of \cite{Pound:2013faa} to obtain the gravitational self-force of the particle. The Weyl scalar $\psi_4$ itself is obtained using an arbitrary precision numerical implementation of the Mano-Suzuki-Tagasugi formalism for solving the Teukolsky equation.

The consistency checks examined in Sec.~\ref{sec:results} provide a great deal of confidence that the numerical implementation of our method is working as expected and is providing accurate results. The accuracy of the results for larger eccentricities currently seems to be limited by the large cancellations in the sum over frequency modes identified in section \ref{sec:TDrecon}. The analysis of that section reveals that these cancellations are inherent to the employed method of extended homogeneous solutions. As the cancellations grow approximately as $(1+e)^{2l}$, they seriously hinder extending the calculation to higher eccentricities and increased accuracies (requiring more $l$-modes). The increased understanding of the cause of these cancellations will, however, assist us in mitigating its consequences in future calculations.

More physics can be extracted from this method by calculating physical observables. One such observable that has previously been calculated for nearly circular orbits in Schwarzschild spacetime \cite{Barack:2011ed} is the periapsis shift. One current obstruction to calculating this quantity is that it is only invariant under a restrict class of gauges. As discussed in \cite{Shah:2015nva}, the ``no-string'' radiation gauge used in this paper is not in the right class of gauges to calculate such pseudo-invariants. The solution offered in \cite{Shah:2015nva} for the Schwarzschild case can be adapted to Kerr, at least for circular orbits. That will be enough to allow calculation of the periapsis shift and the shift of the innermost stable circular equatorial orbit. These calculations will be published in a forthcoming paper \cite{periapsis}. For now we will suffice with noting that preliminary results for the ISCO shift appear to be in perfect agreement with \cite{Isoyama:2014mja}.

The results of the method presented here can in the future also be used to model the evolution of an EMRI around a rotating black hole. This can proceed using osculating geodesic schemes similar to the ones used for the evolution of Schwarzschild inspirals \cite{Warburton:2011fk,Osburn:2015duj}. Depending on the details of the evolution scheme this may again require finding an appropriate gauge part of the completion.  To obtain templates accurate enough to do high precision measurements on future EMRI detections with eLISA, the second order (in $\mr$) correction dissipative part of the self-force will also be needed \cite{Hinderer:2008dm}. Recently some good progress has been made towards obtaining the second order field~\cite{Pound:2015wva}.

The most obvious generalization of our results here is calculation of the GSF for generic inclined orbits around a Kerr black hole. There appear to be no fundamental obstructions to extend the methods used in this paper to such calculation. Doing this will open a new range of physical phenomena to explore such as the shift of the innermost stable spherical orbit (ISSO). It will also allow a direct study of resonances between the radial and polar motion driven by the gravitational self-force first discussed in \cite{Flanagan:2010cd}. In particular it will allow for evaluation of the conditions leading to locking of the resonance \cite{vandeMeent:2013sza}.

\section*{Acknowledgements}
The author wishes to thank Leor Barack, Adam Pound for many useful discussions. He also thanks Sarp Akcay and Niels Warburton for providing the Lorenz gauge data shown in Fig.~\ref{fig:compplot}.
The author was supported by NWO Rubicon grant 680-50-1203 and the European Research Council under the European Union's Seventh Framework Programme (FP7/2007-2013) ERC grant agreement no. 304978. The numerical results in this paper were obtained using the IRIDIS High Performance Computing Facility at the University of Southampton.

\appendix
\section{Conventions}\label{app:A}
\subsection{Background metric}
Through out this paper we work  in ``modified'' Boyer-Lindquist coordinates where the polar coordinate $\theta$ has been replaced by $z=\cos\theta$. In these coordinates the Kerr metric becomes,
\begin{equation}
\begin{split}
\label{eq:kerr}
\id{s}^2 = 
-\bh{1 - \frac{2r}{\Sigma}}\id{t}^2 
+ \frac{\Sigma}{\Delta} \id{r}^2
+ \frac{\Sigma}{1-z^2} \id{z}^2
\\
+ \frac{1-z^2}{\Sigma} \bh{2a^2 r (1-z^2)+(a^2+r^2)\Sigma}\id\phi^2
\\
- \frac{4ar(1-z^2)}{\Sigma}\id{t}\id\phi,
\end{split}
\end{equation}
with
\begin{align}
\Delta &= r(r-2) + a^2,\\
\Sigma &= r^2 + a^2 z^2.
\end{align}
\subsection{Tetrad}
Many of the calculations presented in this paper rely on the Newman-Penrose (NP) formalism. As a null tetrad, we pick the common Kinnersley tetrad expressed in modified Boyer-Lindquist coordinates,
\begin{alignat}{3}
\tet{1}{\mu} &= l^\mu &&= \frac{1}{\Delta}(r^2+a^2,\Delta,0,a),\\
\tet{2}{\mu} &= n^\mu &&= \frac{1}{2\Sigma}(r^2+a^2,-\Delta,0,a),\\
\tet{3}{\mu} &= m^\mu &&= -\frac{\bar\rho \sqrt{1-z^2}}{\sqrt{2}}(\ii a,0,-1,\frac{\ii}{1-z^2}),\\
\tet{4}{\mu} &= \bar{m}^\mu &&= \frac{\rho \sqrt{1-z^2}}{\sqrt{2}}(\ii a,0,1,\frac{\ii}{1-z^2}),
\end{alignat}
with
\begin{equation*}
\rho= \frac{-1}{r-\ii a z}.
\end{equation*}

\subsection{Spin coefficients}
The NP formalism expresses the GR equations in terms of Ricci rotation coefficients
\begin{equation}
\gamma_{abc} \equiv g_{\mu\lambda}\tet{a}{\mu}\tet{c}{\nu}\CD{\nu}\tet{b}{\lambda},
\end{equation}
which are given
\begin{equation}
\begin{aligned}
\kappa &\equiv -\gamma_ {311}, 
	& \varpi &\equiv -\gamma_{241},
		& \epsilon &\equiv -\frac{\gamma_ {211}+\gamma_ {341}}{2}, \\
 \tau &\equiv -\gamma_{312} ,
	& \nu &\equiv -\gamma_ {242}, 	
		& \gamma &\equiv -\frac{\gamma_ {212}+\gamma_ {342}}{2},\\
\sigma &\equiv -\gamma_ {313},
	&	\mu &\equiv -\gamma_{243},
		& \beta &\equiv -\frac{\gamma_{213}+\gamma_ {343}}{2},\\
 \rho &\equiv -\gamma_{314} ,
 	& \lambda &\equiv -\gamma_ {244} ,
 		& \alpha &\equiv -\frac{\gamma_ {214}+\gamma_ {344}}{2} .
\end{aligned}
\end{equation}
Please note the overall sign difference with respect to for example \cite{ChandraBook}.
These signs (and those of other NP quantities) have been chosen such that their background values agree with those common in sources using the NP formalism with a $(+---)$ signature metric (e.g. \cite{ChandraBook,Teukolsky:1972my,Teukolsky:1973ha}). 

For example the Weyl curvature scalars are defined,
\begin{alignat}{3}
\psi_0 &\equiv C_{1313} &&= C_{\mu\nu\rho\sigma}l^\mu m^\nu l^\rho m^\sigma,\\
\psi_1 &\equiv C_{1213} &&= C_{\mu\nu\rho\sigma}l^\mu n^\nu l^\rho m^\sigma,\\
\psi_2 &\equiv C_{1342} &&= C_{\mu\nu\rho\sigma}l^\mu m^\nu \bar{m}^\rho n^\sigma,\\
\psi_3 &\equiv C_{1242} &&= C_{\mu\nu\rho\sigma}l^\mu n^\nu \bar{m}^\rho n^\sigma,\\
\psi_4 &\equiv C_{2424} &&= C_{\mu\nu\rho\sigma}n^\mu \bar{m}^\nu n^\rho \bar{m}^\sigma,
\end{alignat}
where $C_{\mu\nu\rho\sigma}$ is the Weyl tensor.

The directional tetrad derivative operators are defined,
\begin{align}
\Dop &= l^{\mu}\partial_{\mu},\\
\Delop&= n^{\mu}\partial_{\mu},\\
\delop&= m^{\mu}\partial_{\mu},\\
\delopbar&= \bar{m}^{\mu}\partial_{\mu}.
\end{align}
\subsection{Background values}
With these definitions the spin coefficients take the following values on the Kerr background,
\begin{equation}
\kappa=\lambda=\nu=\sigma=\epsilon=0,
\end{equation}
and
\begin{align}
\rho &= \frac{-1}{r- \ii a z},\\
\varpi &= \frac{\ii a \rho^2\sqrt{1-z^2}}{\sqrt{2}},\\
\tau &= -\frac{\ii a \sqrt{1-z^2}}{\Sigma\sqrt{2}},\\
\mu &= \frac{\rho\Delta}{2\Sigma},\\
\gamma &= \frac{\rho\Delta+r-1}{2\Sigma},\\
\beta &= -\frac{\bar\rho z}{2\sqrt{2}\sqrt{1-z^2}},\\
\alpha &= \varpi-\bar\beta.
\end{align}
And the Weyl scalars become
\begin{equation}
\psi_0=\psi_1=\psi_3=\psi_4=0,
\end{equation}
and
\begin{equation}
\psi_2=\rho^3.
\end{equation}
\begin{widetext}
\section{Explicit expressions for self-force contribution from the completion}\label{app:GSFcomp}
To calculate the self-force contribution from the completion, we start by evaluating \eqref{eq:hM} and \eqref{eq:hJ} using \eqref{eq:kerr},
\begin{align}
\begin{split}
h_{\mu\nu}^M &=
	\frac{2 r \left(3 \Sigma -2 r^2\right)}{\Sigma ^2}{\id t}^2
	+\frac{2 r (r (\Delta -\Sigma )+3 \Sigma )}{\Delta ^2}{\id r}^2
	-\frac{2a^2  z^2}{1-z^2}{\id z}^2
	-\frac{8 a^3 r z^2 \hh{1-z^2}}{\Sigma ^2} \id t \id\phi
	\\
	&\quad\quad
	-\Bh{
	\frac{2a^2r\hh{1-z^2}^2\hh{2 r^2-\Sigma}}{\Sigma ^2}
	+2a^2  \hh{1-z^2} 
}{\id\phi}^2,
\end{split}
\\
\intertext{and}
\begin{split}
h_{\mu\nu}^J &=
	-\frac{4 a r z^2}{\Sigma ^2}{\id t}^2
	+\frac{2 a (z^2\Delta -\Sigma )}{\Delta ^2}{\id r}^2
	+\frac{2a z^2}{1-z^2}{\id z}^2
	-\frac{4r \hh{1-z^2}\hh{2 r^2-\Sigma}}{\Sigma ^2} \id t \id\phi
	\\
	&\quad\quad
	+\Bh{
	\frac{4ar^3\hh{1-z^2}^2}{\Sigma^2}
	+2a^2  \hh{1-z^2} 
}{\id\phi}^2.
\end{split}
\end{align}
Evaluating the formula for the self-force \eqref{eq:GSFdef} on these expressions produces the desired $F^{\mu,M/J}_{\mathrm{comp}}$. As expected from symmetry we find  $F^{z,M/J}_{\mathrm{comp}}=0$. The other non-vanishing functions are given by
\begin{align}
\begin{split}
F^{t,M}_{\mathrm{comp}} &= \frac{r_0'}{\Delta _0^4 r_0^5}
\Bh{
\\
&\mathcal{E}^3 \left(a^2 r_0+2 a^2+r_0^3\right) \left(3 a^6+2 a^4 r_0^3+8 a^4 r_0^2-18 a^4 r_0+2 a^2 r_0^5+3 a^2 r_0^4+18 a^2 r_0^3-24 a^2 r_0^2+2 r_0^6+4 r_0^5\right)
\\
&-2 a \mathcal{E}^2 \mathcal{L} \left(4 a^6 r_0+11 a^6+26 a^4 r_0^3+24 a^4 r_0^2-66 a^4 r_0+22 a^2 r_0^5-5 a^2 r_0^4+34 a^2 r_0^3-56 a^2 r_0^2+18 r_0^6-12 r_0^5\right)
\\
&+\mathcal{E} \mathcal{L}^2 \bh{5 a^6 r_0+26 a^6-8 a^4 r_0^4+23 a^4 r_0^3+118 a^4 r_0^2-156 a^4 r_0-8 a^2 r_0^6+31 a^2 r_0^5-26 a^2 r_0^4+76 a^2 r_0^3
\\
&\quad-80 a^2 r_0^2-3 r_0^7+10 r_0^6-8 r_0^5}
-2 a \mathcal{L}^3 \left(5 a^4-8 a^2 r_0^3+34 a^2 r_0^2-30 a^2 r_0-3 r_0^4+10 r_0^3-8 r_0^2\right)
\\
&-\Delta _0 \mathcal{E} \left(2 a^6 r_0+4 a^6+2 a^4 r_0^4+7 a^4 r_0^3-6 a^4 r_0^2-8 a^4 r_0+2 a^2 r_0^6+4 a^2 r_0^5+14 a^2 r_0^4-24 a^2 r_0^3+3 r_0^7\right)
\\
&+2 a \Delta _0 \mathcal{L} \left(2 a^4+6 a^2 r_0^3-a^2 r_0^2-4 a^2 r_0+9 r_0^4-12 r_0^3\right)
},
\end{split}
\\
\begin{split}
F^{r,M}_{\mathrm{comp}} &= \frac{1}{\Delta _0^3 r_0^5}
\Bh{
\\
&\mathcal{E}^4 \left(a^2 r_0+2 a^2+r_0^3\right) \left(3 a^6+2 a^4 r_0^3+8 a^4 r_0^2-18 a^4 r_0+2 a^2 r_0^5+3 a^2 r_0^4+18 a^2 r_0^3-24 a^2 r_0^2+2 r_0^6+4 r_0^5\right)
\\
&-4 a \mathcal{E}^3 \mathcal{L} \left(2 a^6 r_0+7 a^6+14 a^4 r_0^3+16 a^4 r_0^2-42 a^4 r_0+12 a^2 r_0^5-a^2 r_0^4+26 a^2 r_0^3-40 a^2 r_0^2+10 r_0^6-4 r_0^5\right)
\\
&+\mathcal{E}^2 \mathcal{L}^2 \bh{2 a^6 r_0+48 a^6-10 a^4 r_0^4+19 a^4 r_0^3+232 a^4 r_0^2-288 a^4 r_0-10 a^2 r_0^6+32 a^2 r_0^5-38 a^2 r_0^4+200 a^2 r_0^3\\
&\quad-192 a^2 r_0^2-5 r_0^7+10 r_0^6}
+4 a \mathcal{E} \mathcal{L}^3 \left(2 a^4 r_0-9 a^4+18 a^2 r_0^3-66 a^2 r_0^2+54 a^2 r_0+11 r_0^4-34 r_0^3+24 r_0^2\right)
\\
&- \mathcal{L}^4\left(r_0-2\right)\left(5 a^4-8 a^2 r_0^3+34 a^2 r_0^2-30 a^2 r_0-3 r_0^4+10 r_0^3-8 r_0^2\right)
\\
&- \mathcal{E}^2\Delta _0 \left(2 a^6 r_0+4 a^6+2 a^4 r_0^4+7 a^4 r_0^3-6 a^4 r_0^2-8 a^4 r_0+2 a^2 r_0^6+4 a^2 r_0^5+14 a^2 r_0^4-24 a^2 r_0^3+3 r_0^7\right)
\\
&+4 a \mathcal{E} \mathcal{L} \Delta _0 \left(a^4 r_0+2 a^4+7 a^2 r_0^3-6 a^2 r_0^2-4 a^2 r_0+6 r_0^4-6 r_0^3\right)
\\
&- \mathcal{L}^2 \Delta _0\left(2 a^4 r_0+4 a^4-6 a^2 r_0^4+25 a^2 r_0^3-18 a^2 r_0^2-8 a^2 r_0-r_0^5+2 r_0^4\right)
},
\end{split}
\\
\begin{split}
F^{\phi,M}_{\mathrm{comp}} &= \frac{r_0'}{\Delta _0^4 r_0^5}
\Bh{
2 a \mathcal{E}^3 \left(3 a^6+2 a^4 r_0^3+8 a^4 r_0^2-18 a^4 r_0+2 a^2 r_0^5+3 a^2 r_0^4+18 a^2 r_0^3-24 a^2 r_0^2+2 r_0^6+4 r_0^5\right)
\\
&+\mathcal{E}^2 \mathcal{L} \bh{3 a^6 r_0-22 a^6+2 a^4 r_0^4+4 a^4 r_0^3-114 a^4 r_0^2+132 a^4 r_0+2 a^2 r_0^6-a^2 r_0^5+12 a^2 r_0^4-124 a^2 r_0^3
\\
&\quad+112 a^2 r_0^2+2 r_0^7-8 r_0^5}
-2 a \mathcal{E} \mathcal{L}^2 \left(4 a^4 r_0-13 a^4+28 a^2 r_0^3-98 a^2 r_0^2+78 a^2 r_0+19 r_0^4-58 r_0^3+40 r_0^2\right)
\\
&+ \mathcal{L}^3 \left(r_0-2\right)\left(5 a^4-8 a^2 r_0^3+34 a^2 r_0^2-30 a^2 r_0-3 r_0^4+10 r_0^3-8 r_0^2\right)
\\
&-2 a  \mathcal{E}\Delta _0\left(2 a^4 r_0+2 a^4+8 a^2 r_0^3-11 a^2 r_0^2-4 a^2 r_0+3 r_0^4\right)
\\
&+\mathcal{L}\Delta _0  \left(2 a^4 r_0+4 a^4-6 a^2 r_0^4+25 a^2 r_0^3-18 a^2 r_0^2-8 a^2 r_0-r_0^5+2 r_0^4\right)
},
\end{split}
\\
\begin{split}
F^{t,J}_{\mathrm{comp}} &= \frac{r_0'}{\Delta _0^4 r_0^5}
\Bh{
-2 a \mathcal{E}^3 \left(a^2 r_0+2 a^2+r_0^3\right) \left(a^4+a^2 r_0^3+3 a^2 r_0^2-6 a^2 r_0+r_0^5+8 r_0^3-8 r_0^2\right)
\\
&+2 \mathcal{E}^2 \mathcal{L} \left(3 a^6 r_0+8 a^6+21 a^4 r_0^3+20 a^4 r_0^2-48 a^4 r_0+15 a^2 r_0^5-14 a^2 r_0^4+28 a^2 r_0^3-32 a^2 r_0^2-3 r_0^7+10 r_0^6-8 r_0^5\right)
\\
&-4 a \mathcal{E} \mathcal{L}^2 \left(a^4 r_0+5 a^4-2 a^2 r_0^4+4 a^2 r_0^3+24 a^2 r_0^2-30 a^2 r_0-2 r_0^6+7 r_0^5-9 r_0^4+10 r_0^3-8 r_0^2\right)
\\
&+8 a^2 \mathcal{L}^3 \left(a^2-2 r_0^3+7 r_0^2-6 r_0\right)
+2 a \mathcal{E}\Delta _0  \left(2 a^4+a^2 r_0^4+a^2 r_0^3+2 a^2 r_0^2-4 a^2 r_0+r_0^6-r_0^5+10 r_0^4-12 r_0^3\right)
\\
&+2 \mathcal{L}\Delta _0  \left(a^4 r_0-2 a^4-2 a^2 r_0^3-4 a^2 r_0^2+4 a^2 r_0+3 r_0^5-12 r_0^4+12 r_0^3\right)
},
\end{split}
\\
\begin{split}
F^{r,J}_{\mathrm{comp}} &= \frac{1}{\Delta _0^3 r_0^5}
\Bh{
-2 a \mathcal{E}^4 \left(a^2 r_0+2 a^2+r_0^3\right) \left(a^4+a^2 r_0^3+3 a^2 r_0^2-6 a^2 r_0+r_0^5+8 r_0^3-8 r_0^2\right)
\\
&+2 \mathcal{E}^3 \mathcal{L} \left(3 a^6 r_0+10 a^6+23 a^4 r_0^3+26 a^4 r_0^2-60 a^4 r_0+17 a^2 r_0^5-14 a^2 r_0^4+44 a^2 r_0^3-48 a^2 r_0^2-3 r_0^7+10 r_0^6-8 r_0^5\right)
\\
&-2 a \mathcal{E}^2 \mathcal{L}^2 \left(a^4 r_0+18 a^4-5 a^2 r_0^4+7 a^2 r_0^3+92 a^2 r_0^2-108 a^2 r_0-5 r_0^6+16 r_0^5-32 r_0^4+64 r_0^3-48 r_0^2\right)
\\
&-2 \mathcal{E} \mathcal{L}^3 \left(3 a^4 r_0-14 a^4+32 a^2 r_0^3-106 a^2 r_0^2+84 a^2 r_0-3 r_0^5+16 r_0^4-28 r_0^3+16 r_0^2\right)
\\
&+4 a  \mathcal{L}^4 \left(r_0-2\right)\left(a^2-2 r_0^3+7 r_0^2-6 r_0\right)
+2 a  \mathcal{E}^2\Delta _0 \left(2 a^4+a^2 r_0^4+a^2 r_0^3+2 a^2 r_0^2-4 a^2 r_0+r_0^6-r_0^5+10 r_0^4-12 r_0^3\right)
\\
&-4 \mathcal{E} \mathcal{L}\Delta _0  \left(2 a^4+4 a^2 r_0^3-a^2 r_0^2-4 a^2 r_0-2 r_0^5+7 r_0^4-6 r_0^3\right)
+2 a  \mathcal{L}^2 \Delta _0\left(2 a^2-3 r_0^4+9 r_0^3-4 r_0^2-4 r_0\right)
},
\end{split}
\\
\intertext{and}
\begin{split}
F^{\phi,J}_{\mathrm{comp}} &= \frac{r_0'}{\Delta _0^4 r_0^5}
\Bh{
-4 a^2 \mathcal{E}^3 \left(a^4+a^2 r_0^3+3 a^2 r_0^2-6 a^2 r_0+r_0^5+8 r_0^3-8 r_0^2\right)
\\
&-2 a \mathcal{E}^2 \mathcal{L} \left(a^4 r_0-8 a^4+a^2 r_0^4+a^2 r_0^3-44 a^2 r_0^2+48 a^2 r_0+r_0^6-2 r_0^5+14 r_0^4-44 r_0^3+32 r_0^2\right)
\\
&+2 \mathcal{E} \mathcal{L}^2 \left(3 a^4 r_0-10 a^4+24 a^2 r_0^3-78 a^2 r_0^2+60 a^2 r_0-3 r_0^5+16 r_0^4-28 r_0^3+16 r_0^2\right)
\\
&-4 a \mathcal{L}^3 \left(r_0-2\right)  \left(a^2-2 r_0^3+7 r_0^2-6 r_0\right)
+2 \mathcal{E}  \Delta _0 \left(a^4 r_0+2 a^4+6 a^2 r_0^3-6 a^2 r_0^2-4 a^2 r_0-r_0^5+2 r_0^4\right)
\\
&-2 a \mathcal{L} \Delta _0  \left(2 a^2-3 r_0^4+9 r_0^3-4 r_0^2-4 r_0\right)
}.
\end{split}
\end{align}
\end{widetext}

\raggedright
\bibliography{../bib/journalshortnames,../bib/meent,../bib/commongsf,gsfkerr}

\end{document}